%% file: main.tex
\pdfoutput=1
\documentclass[aps,pra,reprint,superscriptaddress,amsmath,amssymb,floatfix,showkeys]{revtex4-2}
\usepackage{array}[=2016-10-06]
\usepackage{imaging_preprint}
\input{results/numerical_values.tex}
\input{results/phase_information_macros.tex}
\input{results/hardware_values.tex}

\input{results/hardware_layout_values.tex}

\begin{document}
\title{Quantum Computing through Computational Imaging:\\
Representations, Measurement, and Inference}
\author{George Kourousias}
\email{george.kourousias@elettra.eu}
\affiliation{Elettra Sincrotrone Trieste, Scientific \& Quantum Computing Unit,
Strada Statale 14--km 163.5, 34149 Trieste, Italy}
\affiliation{University of Trieste, Image Processing Laboratory,
Department of Engineering and Architecture,
Piazzale Europa 1, 34127 Trieste, Italy}
\affiliation{National Centre of Scientific Research ``Demokritos'',
Institute of Quantum Computing and Quantum Technology,
15341 Agia Paraskevi, Athens, Greece}
\author{Sergio Carrato}
\affiliation{University of Trieste, Image Processing Laboratory,
Department of Engineering and Architecture,
Piazzale Europa 1, 34127 Trieste, Italy}
\author{Roberto Pugliese}
\affiliation{Elettra Sincrotrone Trieste, Scientific \& Quantum Computing Unit,
Strada Statale 14--km 163.5, 34149 Trieste, Italy}
\author{Panagiotis Dimitrakis}
\affiliation{National Centre of Scientific Research ``Demokritos'',
Institute of Quantum Computing and Quantum Technology,
15341 Agia Paraskevi, Athens, Greece}
\author{Alberto Carini}
\affiliation{University of Trieste, Image Processing Laboratory,
Department of Engineering and Architecture,
Piazzale Europa 1, 34127 Trieste, Italy}
\date{29 October 2026}
\begin{abstract}
An algebraic correspondence between an imaging model and a quantum state
does not specify a computational method: input access, measurement and
decoding determine what can be returned. This exposition develops four
distinctions for computational-imaging and engineering readers. It
separates amplitude encodings of input values from encodings of complete
candidate labelings, explaining why amplitude storage does not provide
full-array readout. Complex-field interference motivates unknown-phase
inference: complementary measurements resolve ambiguities that local
Fisher information alone does not detect. Reshaping joint amplitudes into
a matrix connects singular-value decomposition to pure-state entanglement,
while distinguishing Schmidt rank from conditions for efficient classical
simulation. In a four-pixel segmentation, an objective becomes a cost
operator and a sampling circuit. One layer of the quantum approximate
optimization algorithm (QAOA) lowers mean energy from
$\QUniformEnergy$ to $\QMeanEnergy$ relative to uniform sampling, yet makes a
nonoptimal labeling more probable than the optimum
($\QRunnerProb$ versus $\QOptProb$). Archived IBM and IQM measurements illustrate
a separate six-variable local boundary decision. Low-rank state
reconstruction and selected quantum algorithms connect these distinctions
to an end-to-end resource account. The derivations and reproducible
numerical references assume complex linear algebra and probability,
without prior quantum mechanics. The accompanying Seeing Amplitudes
browser laboratory supports interactive exploration; the article remains
self-contained and connects each representation to the estimate or
decision it permits and the resources that output requires.
\end{abstract}
\keywords{Computational imaging; quantum computing; quantum measurement;
quantum optimization; statistical inference}
\maketitle

\input{sections/01_introduction.tex}
\input{sections/02_states.tex}

\input{sections/03_transformations.tex}

\input{sections/04_measurement.tex}

\input{sections/05_composites.tex}

\input{sections/07_imaging_example.tex}

\input{sections/06_algorithms.tex}
\input{sections/08_practical.tex}
\section*{Supporting material and data availability}
The supplementary materials are openly available in the article's
\href{https://doi.org/10.34965/enodo.i61162}{Enodo dataset}
\citep{ImagingExpositionDataset}. The numerical materials (S1) contain the
numerical reference, independent validators, figure data, archived
processor counts and native programs, environment records and reproduction
instructions. The web companion materials (S2) are archived in the same
dataset.

The accompanying browser laboratory
\href{https://quantum-works.elettra.eu/seeing-amplitudes/}{\emph{Seeing Amplitudes}}
\citep{seeingamplitudes2026} provides an integrated, interactive route
through the exposition. Readers can change encodings, phases, measurement
settings and shot budgets, inspect state factorizations, and follow an
imaging objective through sampling and decoding. These linked experiments
make the consequences of the article's mathematical choices directly
explorable. The article and numerical reference remain self-contained.
\section*{Acknowledgment}
The authors thank Francesco Guzzi for assessing the suitability of the
material for imaging engineers and for feedback that helped improve the
article. They also thank IBM and IQM for providing free access to their
quantum processors through their cloud platforms. This work received no
dedicated external funding.
\paragraph*{AI assistance.}
The LLMs GPT-6.1 Sol and Claude Opus~5.5 supported selected explanatory
passages on phase inference, matrix factorization and imaging decisions,
as well as revisions, figure refinement, code and consistency checks
(Sections~\ref{sec:measurement},~\ref{sec:entanglement} and~\ref{sec:outlook}).
Opus~5.5 and Elettra's in-house AI also assisted the development of the
web companion. The authors directed all the work, carefully verified the
outputs and remain responsible for all content.
\bibliographystyle{IEEEtran}
\bibliography{references}

\end{document}

%% file: results/numerical_values.tex
\newcommand{\QGamma}{1.9635}
\newcommand{\QBeta}{2.4086}
\newcommand{\QMeanEnergy}{0.686}
\newcommand{\QOptProb}{0.365}
\newcommand{\QRunnerProb}{0.484}

\newcommand{\QUniformEnergy}{1.675}
\newcommand{\QThresholdEnergy}{0.675}
\newcommand{\QOptEnergy}{0.375}
\newcommand{\QShots}{11}
\newcommand{\UniformShots}{72}
\newcommand{\PhaseTruth}{2.30}
\newcommand{\PhaseEstimate}{2.28}
\newcommand{\PhaseReplicateMean}{2.300}
\newcommand{\PhaseReplicateSD}{0.012}
\newcommand{\PhasePredictedSD}{0.011}
\newcommand{\PhaseShots}{4000}
\newcommand{\PhaseReplicates}{1000}

%% file: results/phase_information_macros.tex
\newcommand{\PhaseExampleOverPi}{0.7321127382227185}
\newcommand{\PhaseXYVarianceRatioPlot}{1.012578189297918}

\newcommand{\PhaseDiagnosticCopies}{4000}
\newcommand{\PhaseAllCopies}{12000}

\newcommand{\PhaseXYVarianceRatio}{1.0126}
\newcommand{\PhaseAllCopyVarianceRatio}{1.5189}

%% file: results/hardware_values.tex
\newcommand{\HardwareOptimumMin}{0.346}
\newcommand{\HardwareOptimumMax}{0.611}
\newcommand{\HardwareDecodedMin}{0.695}
\newcommand{\HardwareDecodedMax}{0.781}
\newcommand{\HardwareIdealOptimum}{0.737821}
\newcommand{\HardwareIdealDecoded}{0.801209}
\newcommand{\HardwareIdealOptimumTable}{0.7378}
\newcommand{\HardwareIdealDecodedTable}{0.8012}

%% file: results/hardware_layout_values.tex
\newcommand{\HardwareIBMZeroOptMin}{0.346}
\newcommand{\HardwareIBMZeroOptMax}{0.361}

\newcommand{\HardwareIBMOneOptMin}{0.607}
\newcommand{\HardwareIBMOneOptMax}{0.611}

%% file: sections/01_introduction.tex
\section{Introduction}
\label{sec:thesis}
A computational-imaging researcher routinely manipulates complex arrays,
uses interference to expose phase, factorizes operators and minimizes
objectives. These are useful starting points for understanding quantum
computation. The crucial adjustment concerns \emph{information access}:
a quantum state's amplitudes are not readable array entries, a specified
matrix need not have an inexpensive circuit, and a measured sample need
not be the scientific decision ultimately required.

This article develops four technical distinctions and their consequences,
summarized in Table~\ref{tab:map}. Each is short to state but easy to
misapply when imaging vocabulary is carried over unchanged. We derive
the distinctions from familiar mathematics and follow them through
preparation, measurement, decoding and resource accounting. The result
is a specification of what a proposed computation returns and what a
meaningful comparison must include.

\begin{table*}[!htbp]
\caption{Four distinctions, their development and the examples that make
the consequences explicit.}
\label{tab:map}
\centering\small
\begin{tabular}{@{}>{\raggedright\arraybackslash}p{.17\textwidth}
 >{\raggedright\arraybackslash}p{.32\textwidth}
 >{\raggedright\arraybackslash}p{.14\textwidth}
 >{\raggedright\arraybackslash}p{.27\textwidth}@{}}
\toprule
Distinction & Consequence for an imaging computation & Development & Worked case\\
\midrule
Data versus decision encoding & A basis index can denote a pixel location
or an entire labeling; the amplitudes and returned samples have different meanings.
& Sections~\ref{ssec:encoding}, \ref{ssec:cost_circuit}
& Four values encoded in two qubits versus four binary labels in four qubits.\\[3pt]
Identifiability versus precision & Settings resolve ambiguities;
repetitions quantify uncertainty within the resulting model.
& Sections~\ref{sec:interference}, \ref{ssec:bridge1}, \ref{ssec:phase_information}
& Complementary phase measurements, local information and diagnostic shots.\\[3pt]
Factorization, entanglement and simulation & Singular-value decomposition (SVD) quantifies pure-state
entanglement across a partition; simulation cost also depends on circuit structure.
& Sections~\ref{ssec:coeff}, \ref{sec:exponential}
& Bell state and classical mixture; Schmidt truncation and stabilizer simulation.\\[3pt]
Sampling law versus decoded output & Expected objective, modal assignment
and best sample answer different computational questions.
& Sections~\ref{sec:outlook}, \ref{ssec:hardware_case}
& Four-pixel inference and a local boundary decision on two processors.\\
\bottomrule
\end{tabular}
\end{table*}

The intended reader works in computational imaging, inverse problems,
scientific computing or engineering and knows complex linear algebra,
probability and elementary estimation. No quantum mechanics or specialized
imaging instrument is assumed. The mathematical setting is the standard
circuit model \citep{nielsen_chuang,mermin_qcs,kitaev_shen_vyalyi}.
The exposition explains why amplitude-encoded data are not a readable
array, why local Fisher information can coexist with a global phase
ambiguity, and why Schmidt rank diagnoses pure-state entanglement
without determining simulation difficulty. A complete imaging example
then shows why lowering the mean objective need not make its minimizer
the most probable output. These consequences connect representation,
acquisition and decoding to the useful result of a computation.

Readers new to quantum computing can read
Sections~\ref{sec:state}--\ref{sec:entanglement} in order, then follow
the imaging decision in Section~\ref{sec:outlook}. Readers already
familiar with gates can use Table~\ref{tab:map} to start with
Section~\ref{ssec:encoding}, continue through
Sections~\ref{sec:interference}, \ref{ssec:bridge1}--\ref{ssec:limits}
and~\ref{sec:entanglement}, then read Sections~\ref{sec:outlook}
and~\ref{sec:practical}. Section~\ref{sec:algorithms} gives the broader
algorithmic connections.

\subsection{Relationship to existing accounts}
Engineering-oriented introductions already demonstrate the value of a
disciplinary starting point. Bardin \emph{et al.}\ connect superconducting
quantum computation to microwave engineering
\citep{bardin2020microwave}; Berberich and Fink develop a control
perspective \citep{berberich2024control}; and Au-Yeung \emph{et al.}\
survey algorithms for scientific computing \citep{auyeung2024}.
The imaging correspondences here connect operations to their consequences,
consistent with work on actionable metaphors for quantum computational
thinking \citep{beardow2026metaphors}. Their organizing connection is between an imaging representation,
the measurements that access it, and the estimate or decision returned.
The phase example connects global ambiguity to local Fisher information;
the segmentation example separates mean objective, modal output and
best-sample decoding. These connections let imaging-trained readers assess
what a computation provides before choosing an algorithm or device.

Related applications also distinguish the formulation from its solver.
Jun and Lee formulate joint computed-tomography (CT) reconstruction and
segmentation and use a D-Wave hybrid solver \citep{jun2025segmentation}; this is distinct from
the gate-based sampling circuit developed here. Quantum-native X-ray
fluorescence analysis likewise connects a physical inverse problem to a
discrete objective \citep{kourousias2026xrf}.

The quantum image-processing literature develops dedicated image formats,
including the flexible representation of quantum images (FRQI),
which associates a color-qubit angle with each position, and the novel
enhanced quantum representation (NEQR), which correlates grayscale
basis values with positions
\citep{le2011frqi,zhang2013neqr}. They differ from both encodings in
Table~\ref{tab:encodings}. Our access analysis asks how any such
representation is prepared and which useful output is measured;
it does not impose the arbitrary amplitude-loading cost on every
structured input or image format. Quantum imaging as optical sensing,
including ghost and sub-shot-noise imaging, is outside scope
\citep{pittman1995ghost,bennink2002classicalghost,brida2010imaging}.

\subsection{From preparation to an estimate or decision}
\label{sec:primitives}
In an ideal pure-state description, a register---a collection of qubits
treated jointly---is represented by one normalized complex vector.
One \emph{shot} prepares the register, applies unitary gates and measures
it. With $\ket{\psi_0}$ the prepared state, $U_j$ the gates and $k$ the
classical outcome label,
\begin{equation}
 \ket{\psi_0}\xrightarrow{U_L\cdots U_1}\ket\psi
 \xrightarrow{\text{readout}} k\sim|\psi_k|^2.
 \label{eq:operational_model}
\end{equation}
The left-hand state is the output of preparation; transformation
and measurement follow. Figure~\ref{fig:philosophy} includes
the classical inference or decoding that completes the workflow.
Each new shot invokes preparation again. Variational methods may
additionally update parameters between acquisitions.
Mixed states and noise extend this description in
Section~\ref{sec:entanglement}.
\input{figures/philosophy.tex}
The main calculations use ideal circuits, with independent repeated
preparations for their sampling model; the pixel arrays are synthetic.
Section~\ref{ssec:hardware_case} adds archived IBM/IQM measurements on a
distinct six-pixel task from a related provenance study
\citep[Supplementary Section~S5]{KourousiasProvenance2026}.
These small instances support complete classical checks; they establish
no quantum speedup.

\emph{Seeing Amplitudes} accompanies the article as a browser-based
laboratory \citep{seeingamplitudes2026}. Its routes follow the four
distinctions of Table~\ref{tab:map}: readers compare encodings, choose
phase-measurement settings and shot budgets, inspect state
factorizations, and follow an imaging objective through sampled and
decoded outputs. The displays separate simulator-visible amplitudes
and exact probabilities from simulated samples and archived processor
frequencies. Readers can investigate the consequences of the choices
derived here without installing scientific software. The article
supplies the arguments and numerical references independently;
clearly labeled companion extensions provide further exploration.

%% file: figures/philosophy.tex
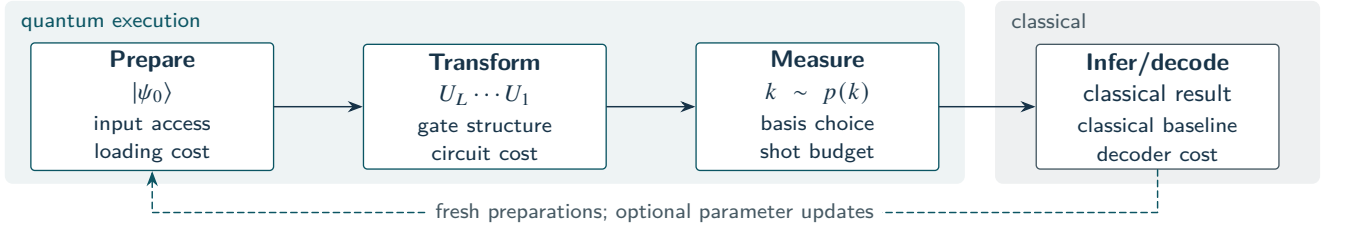
\begin{figure*}[!tp]
  \centering
  \begin{adjustbox}{max width=\linewidth}
\begin{tikzpicture}[
      stage/.style={fig box,draw=FigureTeal,fill=white,text width=30mm,
        minimum height=16.5mm,inner sep=3pt}]
    \fill[FigureTeal!7,rounded corners=3pt] (-1.95,-1.12) rectangle (10.75,1.3);
    \fill[FigureMuted!9,rounded corners=3pt] (11.15,-1.12) rectangle (15.45,1.3);
    \node[anchor=north west,font=\figsmall,text=FigureTeal] at (-1.85,1.25) {quantum execution};
    \node[anchor=north west,font=\figsmall,text=FigureMuted] at (11.25,1.25) {classical};
    \node[stage] (s) at (0,-0.1) {{\figtitle Prepare}\\[1pt]$\ket{\psi_0}$\\[2pt]
      {\figsmall input access\\loading cost}};
    \node[stage] (t) at (4.4,-0.1) {{\figtitle Transform}\\[1pt]$U_L\cdots U_1$\\[2pt]
      {\figsmall gate structure\\circuit cost}};
    \node[stage] (m) at (8.8,-0.1) {{\figtitle Measure}\\[1pt]$k\sim p(k)$\\[2pt]
      {\figsmall basis choice\\shot budget}};
    \node[stage,draw=FigureMuted] (c) at (13.3,-0.1) {{\figtitle Infer/decode}\\[1pt]classical result\\[2pt]
      {\figsmall classical baseline\\decoder cost}};
    \draw[fig arrow] (s) -- (t);
    \draw[fig arrow] (t) -- (m);
    \draw[fig arrow] (m) -- (c);
    \draw[fig arrow,draw=FigureTeal,dash pattern=on 2.4pt off 1.6pt] (c.south) -- (13.3,-1.5)
      -- node[fill=white,inner sep=3pt,font=\figsmall,text=FigureMuted]{fresh preparations; optional parameter updates}
      (0,-1.5) -- (s.south);
  \end{tikzpicture}
\end{adjustbox}
  \caption{One shot proceeds from preparation to readout; classical
    inference combines outcomes into an estimate or decision. Each stage specifies its access or resource
    cost; $p(k)$ depends on the preparation, circuit and measurement
    basis. The dashed return denotes new preparations for further
    acquisition, with optional classical parameter updates.}
  \label{fig:philosophy}
\end{figure*}

%% file: sections/02_states.tex
\section{Representations and information access}
\label{sec:state}

A representation specifies a quantum state and connects its possible
measurement outcomes to the imaging task.

\subsection{One qubit: complex amplitudes and relative phase}

A qubit has two distinguished computational-basis states, whose coordinate
columns are $\ket{0}=(1,0)^{\top}$ and $\ket{1}=(0,1)^{\top}$.
A normalized vector describes a \emph{pure state}:
\begin{equation}
  \begin{aligned}
  \ket{\psi}&=\alpha\ket{0}+\beta\ket{1},\qquad
  \vec{\psi}=\begin{pmatrix}\alpha\\\beta\end{pmatrix},\\
  |\alpha|^2+|\beta|^2&=1.
  \end{aligned}
  \label{eq:qubit}
\end{equation}
The ket $\ket{\psi}$ and column $\vec\psi$ denote the same vector in
this basis. The bra $\bra{\psi}$ is the row $\vec\psi^{\dagger}$
obtained by conjugate transposition.
The complex coefficients $\alpha$ and $\beta$ are \emph{amplitudes}:
each has a magnitude and a phase, as does a phasor or a Fourier
coefficient. A computational-basis measurement returns a classical label
$0$ or $1$, with probabilities
\begin{equation}
  \Pr(k)=|\psi_k|^2.
  \label{eq:born}
\end{equation}
Here $\psi_0=\alpha$ and $\psi_1=\beta$, so
$\Pr(0)=|\alpha|^2$ and $\Pr(1)=|\beta|^2$. The labels $0,1$ name
outcomes; the kets $\ket0,\ket1$ name basis states. This is the Born rule.
The same rule applies to several qubits;
Section~\ref{sec:measurement} develops its statistical and physical
consequences. The normalization in Eq.~\eqref{eq:qubit} makes the
probabilities sum to one.

Multiplying every amplitude by $e^{\ii\chi}$ changes no observable
prediction: the normalized vectors $\ket{\psi}$ and
$e^{\ii\chi}\ket{\psi}$ represent the same
pure state. This \emph{global phase} is distinct from a \emph{relative
phase}. A \emph{superposition} has more than one nonzero coefficient
in the specified basis.
For example,
\begin{equation}
  \ket{+}=\frac{\ket{0}+\ket{1}}{\sqrt2},\qquad
  \ket{-}=\frac{\ket{0}-\ket{1}}{\sqrt2}
\end{equation}
have identical computational-basis probabilities but are different
states. The Hadamard transformation introduced in
Section~\ref{sec:transform} maps them to $\ket0$ and $\ket1$, respectively,
so a subsequent measurement distinguishes them deterministically.
Coherent phase relationships can therefore affect later observations;
a superposition is not merely uncertainty about an unobserved binary value.
The useful imaging comparison is addition of coherent \emph{fields}
rather than addition of incoherent \emph{intensities}. Superposition is
defined relative to a basis; neither $\ket{+}$ nor $\ket{-}$ is a
superposition in the orthonormal basis $\{\ket{+},\ket{-}\}$.

\subsection{Registers, tensor products and indexing}

A \emph{register} is a collection of $n$ qubits treated jointly. Its
pure-state description has $2^n$ entries,
\begin{equation}
  \ket{\psi}=\sum_{k=0}^{2^n-1}\psi_k\ket{k},\qquad
  \sum_k|\psi_k|^2=1.
\end{equation}
Here the integer index $k$ has binary digits labeling the qubits.
Throughout this article the leftmost qubit is most significant:
$\ket{q_1\cdots q_n}$ has array index $\sum_{j=1}^{n}q_j2^{n-j}$, where
$q_j\in\{0,1\}$ is the label of qubit $j$. For example, $\ket{101}$
has index $5$. This convention is part of the computational
specification, not a property of quantum mechanics; software packages
may use different displayed bit orders.

For separately prepared qubits with state vectors
$\ket a=(a_0,a_1)^{\top}$ and $\ket b=(b_0,b_1)^{\top}$, the joint state
is their \emph{tensor product}:
\begin{equation}
  \ket{a}\otimes\ket{b}
   =(a_0b_0,\,a_0b_1,\,a_1b_0,\,a_1b_1)^{\top}.
  \label{eq:kron}
\end{equation}
The operation is \verb|np.kron(a, b)| in a state-vector simulator. As
with separable image operators, a factored representation can be much
smaller than its expanded array. A product of $n$ single-qubit states is
specified by $n$ two-entry vectors. Not every joint pure state admits
this factorization; its failure defines entanglement, discussed in
Section~\ref{sec:entanglement}.

\subsection{An encoding specifies what the state means}
\label{ssec:encoding}

Consider four input values $f_0,\ldots,f_3$. An \emph{amplitude
encoding} assigns them to a two-qubit state,
\begin{equation}
 \begin{aligned}
  \ket{f}&=\frac{1}{\|\vec f\|_2}\sum_{k=0}^{3}f_k\ket{k},\qquad\vec f\ne0,\\
  \ket{k}&\in\{\ket{00},\ket{01},\ket{10},\ket{11}\}.
 \end{aligned}
  \label{eq:ampencoding}
\end{equation}
The binary basis labels $00,01,10,11$ identify locations
$k=0,1,2,3$, respectively; the amplitude $f_k/\|\vec f\|_2$ is the
normalized input value at that location. Sampling in the computational basis returns location $k$ with
probability $|f_k|^2/\|\vec f\|_2^2$; it does not return $f_k$. The
original scale $\|\vec f\|_2$ must be retained separately if needed, and
an overall complex phase cannot be recovered from this state alone.

For nonnegative intensities $g_k$, encoding the \emph{intensity values}
$f_k=g_k$ produces probabilities proportional to $g_k^2$. For
\emph{intensity-proportional sampling}, one instead prepares
$f_k=\sqrt{g_k}$, giving $\Pr(k)=g_k/\sum_j g_j$.
A \emph{coherent-field encoding},
$f_k=\sqrt{g_k}e^{\ii\theta_k}$, also specifies the field phase
$\theta_k$ at location $k$. That relative-phase information is absent
from the intensity array alone. These choices change the state and its
readout law; describing all three as an ``image in a quantum register''
would hide the difference.

An imaging decision problem uses another representation. Suppose each
of four pixels receives a binary label $x_i\in\{0,1\}$. The four decision
qubits have sixteen basis states, one per complete labeling
$\ket{x_1x_2x_3x_4}$, such as $\ket{1101}$. A quantum optimization algorithm
prepares
\begin{equation}
  \ket{\psi}=\sum_{\vec x\in\{0,1\}^{4}}a_{\vec x}\ket{\vec x}.
  \label{eq:decisionencoding}
\end{equation}
The complex search amplitude $a_{\vec x}$ gives probability
$|a_{\vec x}|^2$ for the complete candidate labeling $\vec x$.
It does not encode the intensity of pixel $i$. Sampling returns one
labeling, which a classical decoder evaluates. The observed image may determine circuit parameters
or a cost operator without ever being amplitude-encoded. This is the
representation used in Section~\ref{sec:outlook}.
Table~\ref{tab:encodings} contrasts these two encodings and their readout.
The companion's \href{https://quantum-works.elettra.eu/seeing-amplitudes/routes/representation/\#x-encoding}{encoding experiment}
shows how changing the same four input values changes the readout laws
under these representations.

\begin{table*}[!htbp]
\centering\small
\caption{Two encodings of four-entry imaging problems. Qubit count alone
does not specify either the task or the information returned.}
\label{tab:encodings}
\begin{tabular}{@{}>{\raggedright\arraybackslash}p{3.0cm}
                  >{\raggedright\arraybackslash}p{5.1cm}
                  >{\raggedright\arraybackslash}p{5.1cm}@{}}
\toprule
 & \textbf{Amplitude encoding} & \textbf{Binary-decision encoding}\\
\midrule
Register & Two qubits; four basis states & Four qubits; sixteen basis states\\
Meaning of a basis label & Entry or location $k$ & Complete labeling $\vec x$\\
Meaning of coefficients & Normalized field or chosen data representation & Search amplitudes over labelings\\
Input requirement & Prepare $\ket{f}$ & Encode the objective; prepare an initial search state\\
Computational-basis output & Location sampled with probability proportional to $|f_k|^2$ & Candidate labeling sampled with probability $|a_{\vec x}|^2$\\
\bottomrule
\end{tabular}
\end{table*}

Preparing an arbitrary $N$-entry pure state in the amplitude encoding of
Eq.~\eqref{eq:ampencoding}
generally uses $O(N)$ elementary gates, although the register uses
only $\lceil\log_2N\rceil$ qubits \citep{plesch2011}. Efficient input
preparation depends on structure or a stated access model; extra entries
can be zero-padded when $N$ is not a power of two. The decision encoding
has different costs: preparing a uniform search state is easy, but the
objective's interactions still have to be implemented.

%% file: sections/03_transformations.tex
\section{Unitary transformations and interference}
\label{sec:transform}

We now connect state representations to implementable transformations,
then use interference to make relative phase observable.

\subsection{Norm preservation and circuit cost}

In the ideal closed-system circuit model, a transformation acts as
\begin{equation}
  \vec\psi\longmapsto U\vec\psi,\qquad U^{\dagger}U=UU^{\dagger}=I.
  \label{eq:unitary}
\end{equation}
Here $\dagger$ denotes conjugate transpose and $I$ the identity matrix.
The matrix $U$ is unitary. It preserves norms and inner products and has
inverse $U^{\dagger}$. All its singular values are one: it is a perfectly
conditioned linear transformation. A dense state-vector simulator stores
the amplitude array explicitly; applying the matrix $U$ gives the update
\verb|psi = U @ psi|, where \verb|@| denotes NumPy matrix multiplication.
Normalized Fourier and orthogonal
transforms provide mathematical examples already familiar from imaging
\citep{goodman2005}.

Many useful image-processing operations are not unitary. For example, thresholding
and downsampling discard information. A blur may be invertible on a
chosen finite discretization yet amplify noise severely when inverted;
its singular values are not all one. Such operations cannot simply be
declared gates on the same register. They can sometimes be embedded in
a larger unitary process using auxiliary qubits, followed by measurement
or postselection---retaining only selected outcomes---with the extra resources and success probability
included \citep{nielsen_chuang}.

Unitarity means \emph{norm preservation}, not general conservation of
physical energy. A driven $X$ gate (the bit flip exchanging $\ket0$
and $\ket1$) can change the qubit's physical energy. The optical analogy
concerns redistribution of coherent amplitudes without loss, rather than
the energy budget of every quantum device.

A second distinction concerns implementation. A unitary matrix is
mathematically admissible without being computationally inexpensive.
A generic $n$-qubit unitary has exponentially many independent parameters
and generally needs an exponentially large gate decomposition. Algorithms
use \emph{structured} transformations specified by manageable circuits.
These circuits combine gates acting on one or a few qubits; for example,
a single-qubit $2\times2$ unitary $G$ acts on the first of two qubits
as $G\otimes I_2$, with $I_2$ the $2\times2$ identity. Its full
matrix is not physically constructed or stored. A circuit specification
records local operations, their order and their dependencies.

Circuit diagrams will be read from left to right, as illustrated in
Figure~\ref{fig:hphaseh}. Each horizontal wire
denotes one qubit; boxes are gates, connected dots indicate control and
target operations, and measurement symbols terminate the quantum part
of the calculation. Matrix products act in the opposite written order:
applying $G_1$ first and $G_2$ second produces $G_2(G_1\ket\psi)$. Compilation
later translates these logical gates into a device's supported operations
and connectivity, potentially adding gates and depth. Thus matrix size,
logical circuit cost and physical implementation cost are distinct.

\subsection{Gates as structured operations}

The following single-qubit matrices suffice for the central derivations:
\begin{equation}
 \begin{aligned}
  X&=\begin{pmatrix}0&1\\1&0\end{pmatrix},\quad
  P(\phi)=\begin{pmatrix}1&0\\0&e^{\ii\phi}\end{pmatrix},\\
  H&=\frac{1}{\sqrt2}\begin{pmatrix}1&1\\1&-1\end{pmatrix}.
 \end{aligned}
  \label{eq:gates}
\end{equation}
The $X$ gate swaps amplitudes. The phase gate $P(\phi)$ changes their
relative phase, leaving computational-basis probabilities unchanged.
Writing $Z=P(\pi)=\operatorname{diag}(1,-1)$ and
$S=P(\pi/2)=\operatorname{diag}(1,\ii)$ establishes notation used below.
The factor $1/\sqrt2$ normalizes the orthogonal columns of $H$.
Its action on one qubit's two amplitudes is
$H(\alpha,\beta)^{\top}=(\alpha+\beta,\alpha-\beta)^{\top}/\sqrt2$:
it forms sum and difference channels, exactly as the normalized
two-point discrete Fourier transform (DFT) does. It is a coherent splitter:
$H\ket0=\ket+$, and $H^2=I$ recovers the input exactly.

The matrix exponential is $e^A=\sum_{r=0}^{\infty}A^r/r!$, with
powers denoting matrix products.
Parameterized gates are often specified by an adjustable rotation angle
$\theta$,
\begin{equation}
  R_X(\theta)=e^{-\ii\theta X/2},\qquad
  R_Z(\theta)=e^{-\ii\theta Z/2}.
  \label{eq:rotations}
\end{equation}
For $\sigma\in\{X,Z\}$, $\sigma^2=I$ gives
$e^{-\ii\theta\sigma/2}=\cos(\theta/2)I-\ii\sin(\theta/2)\sigma$.
Consequently, $e^{-\ii aZ}=R_Z(2a)$, while
$P(\phi)=e^{\ii\phi/2}R_Z(\phi)$. Keeping the half-angle convention
explicit avoids a factor-of-two error when turning a Hamiltonian into
software instructions. A Hamiltonian is a Hermitian generator of unitary
evolution; a cost Hamiltonian encodes objective values in the
computational basis. Equivalence up to global phase applies to the
uncontrolled operation: the factor $e^{\ii\phi/2}$ affects only the
control-$\ket1$ branch of controlled-$P(\phi)$, relative to controlled-$R_Z(\phi)$,
and leaves the control-$\ket0$ branch unchanged. It is therefore a relative
phase between branches.
The controlled-\textnormal{NOT} (CNOT) gate acts as
$\ket{a,b}\mapsto\ket{a,b\mathbin\oplus a}$, where $\oplus$ denotes
addition modulo two. It is a reversible interaction between wires and
can create the nonseparable states studied in
Section~\ref{sec:entanglement}.

\subsection{The H--phase--H interference experiment}
\label{sec:interference}

In the single-qubit circuit of Figure~\ref{fig:hphaseh}, prepare $\ket0$,
apply $H$, write a phase with $P(\phi)$, and apply $H$ again.
Direct multiplication gives
\begin{align}
 \ket0&\xrightarrow{H}\frac{\ket0+\ket1}{\sqrt2}
 \xrightarrow{P(\phi)}\frac{\ket0+e^{\ii\phi}\ket1}{\sqrt2}
 \notag\\
 &\xrightarrow{H}\frac{(1+e^{\ii\phi})\ket0+(1-e^{\ii\phi})\ket1}{2}\notag\\
 &=e^{\ii\phi/2}\left(\cos\frac\phi2\ket0
                   -\ii\sin\frac\phi2\ket1\right).
 \label{eq:hph}
\end{align}
The common factor $e^{\ii\phi/2}$ multiplies both components and is
a global phase. Equation~\eqref{eq:born} therefore
gives
\begin{equation}
 \begin{aligned}
  \Pr(0)&=\frac{1+\cos\phi}{2}=\cos^2\frac\phi2,\\
  \Pr(1)&=\frac{1-\cos\phi}{2}=\sin^2\frac\phi2.
 \end{aligned}
  \label{eq:p1}
\end{equation}
Before the second $H$, both outcomes have probability $1/2$ for every
$\phi$. Afterwards they depend on $\phi$. Phase information was present
but unobservable in the first measurement arrangement; recombination
made it observable.

This is the same interference algebra as
\begin{equation}
  |A_1+A_2|^2=|A_1|^2+|A_2|^2
                 +2\operatorname{Re}(A_1^*A_2).
\end{equation}
The cross term carries relative phase. An imaging reader can therefore
derive the fringe rather than treat it as an unfamiliar quantum rule.
The analogy is algebraic: a qubit need not be implemented as two optical
paths.

\input{figures/interferometer.tex}

\noindent\begin{minipage}{\linewidth}
The corresponding ideal state-vector simulation is short:
\begin{lstlisting}
import numpy as np
rng = np.random.default_rng(7)
H = np.array([[1, 1], [1, -1]], complex)
H /= np.sqrt(2)
phi = 2.30
P = np.diag([1, np.exp(1j * phi)])
initial = np.array([1, 0], complex)
psi = H @ P @ H @ initial
p = np.abs(psi)**2
# Sample one freshly prepared copy.
k = rng.choice(2, p=p)
# Conditional projective update.
psi_after = np.eye(2, dtype=complex)[k]
\end{lstlisting}
\end{minipage}
The line computing \verb|p| evaluates exact probabilities for this model;
\verb|rng.choice| draws one simulated outcome from them. Repeated hardware
shots estimate those probabilities through outcome frequencies. Sampling
and the conditional state update are separate operations in this code.
Each repetition starts with a fresh preparation, rather than continuing
the circuit from \verb|psi_after|. This distinction connects the transform
calculation to the inference problem next.

%% file: figures/interferometer.tex
\begin{compactfigure}
\centering
\begin{adjustbox}{max width=\linewidth}
\begin{tikzpicture}
  \begin{scope}[shift={(0.95,0)}]
  \draw[qwire] (0,0) -- (5.2,0);
  \draw[cwire] (5.5,0) -- (6.05,0);
  \node[anchor=east] at (0,0) {$\ket0$};
  \node[gate] at (0.8,0) {$H$};
  \node[gate] at (2.25,0) {$P(\phi)$};
  \node[gate] at (3.7,0) {$H$};
  \pic at (5.2,0) {meter};
  \foreach \x/\t in {0.8/split,2.25/phase,3.7/mix,5.2/readout}{
    \node[font=\figsmall,text=FigureMuted] at (\x,-0.5) {\t};}
  \end{scope}
  \begin{axis}[at={(0.25cm,-1.5cm)},anchor=north west,width=7.9cm,height=3.5cm,
   domain=0:pi,samples=80,xmin=0,xmax=3.3,ymin=0,ymax=1.12,
   xtick={0,1.5708,3.1416},xticklabels={$0$,$\pi/2$,$\pi$},
   ytick={0.5,1},extra y ticks={0},extra y tick labels={},
   axis lines=left,axis line style={FigureMuted,-},tick style={FigureMuted},
   xlabel={Phase $\phi$},ylabel={Probability $\Pr(1)$},
   ylabel style={rotate=-90,at={(axis description cs:0,1.02)},anchor=south west,xshift=-4mm},
   clip=false]
   \addplot[FigureTeal,line width=1.1pt] {sin(deg(x/2))^2};
  \end{axis}
\end{tikzpicture}
\end{adjustbox}
\caption{A single-qubit circuit and its ideal interference fringe.
The phase gate alone leaves both probabilities at $1/2$; the final
Hadamard converts relative phase into outcome contrast. The continuous
curve is a probability, estimated experimentally through repeated shots.}
\label{fig:hphaseh}
\end{compactfigure}

%% file: sections/04_measurement.tex
\section{Measurement and statistical inference}
\label{sec:measurement}

This section develops measurement as a statistical inference problem.

\subsection{Outcomes, bases and back-action}
\label{ssec:born}

One computational-basis shot returns a bitstring, or its equivalent
integer index $k$, with the probability in Eq.~\eqref{eq:born}.
Counts from repeated shots estimate these probabilities. An ideal
projective measurement also changes that prepared copy to $\ket{k}$.
This postmeasurement state generally differs from the original
superposition; an input already equal to $\ket{k}$ is unchanged.
Measuring the same updated copy again therefore does not supply an
independent sample of the original distribution.

For an orthonormal measurement setting labeled $B$, apply an analysis
unitary $V_B$ before computational-basis readout:
\begin{equation}
  p_B(k)=|\braket{k|V_B|\psi}|^2=|(V_B\vec\psi)_k|^2.
  \label{eq:basisreadout}
\end{equation}
The measured basis vectors are $V_B^{\dagger}\ket{k}$: the applied
unitary is the inverse of the matrix whose columns list that basis.
The setting label $B$ and the applied matrix $V_B$ are distinct.
Choosing a setting resembles choosing an image or Fourier intensity
detection plane. Such optical detection does not directly reveal the
complex field; here each quantum shot supplies one sampled index.

When only part of a register is measured, write the joint pure state as
$\sum_a\ket{a}\otimes\ket{\chi_a}$ in the measured subsystem's basis.
Outcome $a$ has probability $\|\chi_a\|^2$ and leaves the unmeasured
subsystem in $\ket{\chi_a}/\|\chi_a\|$, provided this probability is
nonzero. The unmeasured subsystem generally remains a superposition.
This conditional update is essential when an auxiliary qubit, called an
\emph{ancilla}, is used to read information about a larger register.

\subsection{Estimating a quantity rather than reconstructing an array}

For $N$ independent, identically prepared shots in a fixed basis, a
frequency $\widehat p=n_k/N$ has variance $p(1-p)/N$. Statistical error
thus decreases as $N^{-1/2}$. More repetitions buy precision, just as
accumulating independent photon counts does, although fixed-shot
frequencies are binomial or multinomial rather than independent Poisson
counts. This empirical scaling concerns the stated fixed-probe acquisition;
Section~\ref{sec:algorithms} also considers coherent phase accumulation.

Often only an expectation is required. An \emph{observable} is a
Hermitian matrix $A=A^\dagger$ specifying a measured quantity: its
eigenvalues $a_k$ are the possible measured values. It describes the
measurement, not the prepared state. When its eigenbasis can be measured,
the expectation $\langle A\rangle$ and its sample estimator are
\begin{equation}
  \langle A\rangle=\braket{\psi|A|\psi}
       =\sum_k a_kp_A(k),\qquad
  \widehat{\langle A\rangle}=\frac{1}{N}\sum_{r=1}^{N}a_{k_r}.
  \label{eq:observable}
\end{equation}
Here $p_A(k)$ labels the eigenbasis probabilities of the observable $A$.
The estimator variance is
$(\langle A^2\rangle-\langle A\rangle^2)/N$. The Pauli observables
$X$, $Y=\left(\begin{smallmatrix}0&-\ii\\\ii&0\end{smallmatrix}\right)$
and $Z$ have eigenvalues $+1$ and $-1$: bits $0$ and $1$ correspond
to those values after the appropriate analysis gates. A \emph{Pauli
string} is a tensor product of factors from $I,X,Y,Z$. Its measured
value is the product of the nonidentity factors' eigenvalues. If
their measured bits are $b_i$, the result is
$\prod_i(1-2b_i)=(-1)^{\sum_i b_i}$. Thus $Z_iZ_j$ has eigenvalue
$+1$ when the two computational-basis bits agree and $-1$ when they
differ. For
$A=\sum_j c_jP_j$, expectations of Pauli strings $P_j$ can be estimated
and combined classically; incompatible strings generally require
different acquisition settings. The number of settings, shot allocation
and covariance between jointly estimated terms contribute to the cost.

This is an imaging-relevant distinction. Estimating a specified overlap,
energy or other scalar functional may be useful without recovering an
entire state. Conversely, reconstructing a generic $n$-qubit state
requires resolving exponentially many parameters. Tomography and
classical-shadow methods infer properties from designed measurements,
with guarantees depending on the observable family, precision and access
assumptions \citep{nielsen_chuang,huang2020shadows}.

\subsection{Worked phase inference by measurement diversity}
\label{ssec:bridge1}

Consider a repeatable source producing
\begin{equation}
  \ket{\psi_\phi}=\frac{\ket0+e^{\ii\phi}\ket1}{\sqrt2}.
  \label{eq:psiphi}
\end{equation}
The phase $\phi$ is an unknown property of a stable, repeatable source.
The experimenter chooses the gates applied \emph{after} each source
preparation, as in Figure~\ref{fig:phase_diversity}. These gates select
the measurement setting without changing the source's unknown parameter.
Independent preparations are assumed; they do not require cloning an
arbitrary unknown specimen.

Computational ($Z$) readout yields $p_Z(0)=1/2$ for every $\phi$ and hence
contains no phase information. Applying $H$ before readout measures $X$,
revealing $\cos\phi$. Applying $S^{\dagger}=P(-\pi/2)$ and then $H$
measures $Y$, revealing $\sin\phi$. Thus the analysis matrices in
Eq.~\eqref{eq:basisreadout} are $V_Z=I$, $V_X=H$ and $V_Y=HS^\dagger$.
Writing hats for estimates from measured frequencies gives
\begin{equation}
 \begin{aligned}
  p_X(0)&=\frac{1+\cos\phi}{2},\qquad
  p_Y(0)=\frac{1+\sin\phi}{2},\\
  \widehat\phi&=\operatorname{atan2}
       \big(2\widehat p_Y(0)-1,\,2\widehat p_X(0)-1\big).
 \end{aligned}
  \label{eq:xy}
\end{equation}
The phase estimate is understood modulo $2\pi$. An $X$ record alone
leaves $\phi$ and $-\phi$ ambiguous; a $Y$ record resolves this
conjugation ambiguity. Measurement diversity supplies information absent
from a single intensity record, as in phase retrieval, but here the
unknown is the single parameter $\phi$ rather than an entire field.

\input{figures/phase_settings.tex}

With $N$ independent shots \emph{per setting}, the empirical estimators
$\widehat c=2\widehat p_X(0)-1$ and
$\widehat s=2\widehat p_Y(0)-1$ estimate $\cos\phi$ and $\sin\phi$.
Their variances are
$\sin^2\phi/N$ and $\cos^2\phi/N$, respectively. Linearizing
$\operatorname{atan2}(s,c)$ gives the delta-method prediction
\begin{equation}
  \operatorname{Var}(\widehat\phi)
       \simeq\frac{\sin^4\phi+\cos^4\phi}{N}.
  \label{eq:phasevariance}
\end{equation}
The numerical acquisition uses $\phi=\PhaseTruth$, $N=\PhaseShots$
shots in each of $X$ and $Y$, and $\PhaseDiagnosticCopies$ additional
diagnostic $Z$ shots: $\PhaseAllCopies$ preparations in total.
The predicted standard deviation is approximately $\PhasePredictedSD$ rad. The seeded ideal simulations in
the supplement give one estimate of $\PhaseEstimate$ rad and a mean and standard
deviation of $\PhaseReplicateMean\pm\PhaseReplicateSD$ rad over $\PhaseReplicates$ repeated experiments. The
latter is the spread of individual estimates, not an uncertainty of
$\PhaseReplicateSD$ rad on their average. Circular statistics or an appropriate
unwrapping convention are needed near the estimator's branch cut.

\subsection{Local information and global identifiability}
\label{ssec:phase_information}

The same source phase $\phi$ can affect nearby probabilities strongly
even when distant parameter values remain indistinguishable.
Classical Fisher information measures this local sensitivity
\citep{kay1993estimation}. For setting $B\in\{X,Y,Z\}$ and
outcome $k\in\{0,1\}$, write $p_B(k\mid\phi)$ to display the dependence
on $\phi$. The information per shot is
\begin{equation}
 F_B(\phi)=\sum_{k:p_B(k\mid\phi)>0}
 \frac{[\partial_\phi p_B(k\mid\phi)]^2}{p_B(k\mid\phi)}.
 \label{eq:phase_classical_fisher}
\end{equation}
Here $\partial_\phi$ denotes differentiation with respect to $\phi$;
the squared probability changes are weighted by the inverse outcome
probability. Away from phases with deterministic $X$ or $Y$ outcomes, the probabilities
in Eq.~\eqref{eq:xy} give $F_X=F_Y=1$, whereas $F_Z=0$.
An $X$ ensemble therefore contains local information even though it
cannot distinguish $\phi$ from $-\phi$ globally.

For independent ensembles containing $N$ shots in each of the $X$ and
$Y$ settings, information adds to $2N$. The local Cram\'er--Rao variance
bound \citep{kay1993estimation} is consequently $1/(2N)$ for regular unbiased estimation; the
same benchmark governs suitable locally asymptotically unbiased
estimators. The pure-state quantum Fisher information is
\begin{equation}
 F_Q=4\bigl(\langle\partial_\phi\psi_\phi|
 \partial_\phi\psi_\phi\rangle
 -|\langle\psi_\phi|\partial_\phi\psi_\phi\rangle|^2\bigr)=1.
 \label{eq:phase_quantum_fisher}
\end{equation}
This is the measurement-optimized local information per copy for the
stated state family \cite{braunstein1994geometry}. At regular phases the two settings therefore retain the available local
information, although a particular estimator need not use it efficiently.

Combining the bound with Eq.~\eqref{eq:phasevariance} gives the
large-sample variance ratio
\begin{equation}
 R(\phi)\simeq 2(\sin^4\phi+\cos^4\phi),
 \qquad 1\leq R(\phi)\leq2.
 \label{eq:phase_efficiency}
\end{equation}
The simple $\operatorname{atan2}$ estimator therefore trades convenience
for phase-dependent local efficiency. At $\phi=\PhaseTruth$, its predicted
variance is $\PhaseXYVarianceRatio$ times the informative-ensemble
benchmark. This large-sample comparison varies with phase, as
Figure~\ref{fig:phase_efficiency} shows.

The $\PhaseDiagnosticCopies$ diagnostic Z shots check computational-basis
population balance. They do not test coherence: complete dephasing also
leaves $p_Z(0)=1/2$. All $\PhaseAllCopies$ shots belong in the acquisition
account. If every copy were devoted to phase estimation, the
measurement-optimized local benchmark would instead be $1/(3N)$;
the predicted variance has ratio $\PhaseAllCopyVarianceRatio$ to that
benchmark. The $\PhaseXYVarianceRatio$ comparison prices only the
informative X/Y ensembles.
\input{figures/phase_efficiency.tex}

\subsection{Ancillary interferometry and the Hadamard test}
\label{ssec:recovery}

Holography recombines reference and object fields to expose relative
phase \citep{gabor1948}. In the Hadamard test, an ancilla supplies two
coherent branches: the object register is unchanged in one branch and
transformed by a unitary $U$ in the other. Prepare the ancilla and object
in $\ket+\ket\psi$, and apply $U$ only when the ancilla is in $\ket1$:
\begin{equation}
 \ket\Omega=\frac{\ket0\ket\psi+\ket1U\ket\psi}{\sqrt2}.
\end{equation}
Applying $H$ to the ancilla makes the object vector associated with
ancilla outcome $0$ equal to $(\ket\psi+U\ket\psi)/2$. Its squared
norm gives the probability of the ancilla readout bit $a$ being zero:
\begin{equation}
 \Pr(a=0)=\frac{1+\operatorname{Re}\braket{\psi|U|\psi}}{2}.
 \label{eq:hadtest}
\end{equation}
With $S^{\dagger}$ before that final $H$, the corresponding expression
contains the imaginary part. The complex overlap
$\braket{\psi|U|\psi}$ is the inner product between the original and
transformed object states. The two settings estimate this scalar. If
$U\ket\psi=e^{2\pi\ii\theta}\ket\psi$, the real-part readout becomes
$\cos^2(\pi\theta)$, recovering the earlier interference fringe.
Controlled powers of $U$ extend this construction to quantum phase
estimation in Section~\ref{sec:algorithms}.

The final Hadamard recombines the two branches, exposing their overlap
in the ancilla statistics. This requires an implementable controlled-$U$
circuit, repeated preparations for the readout, and the conditional
object-state update described in Section~\ref{ssec:born}. Access to $U$
alone does not guarantee an inexpensive controlled-$U$.

\subsection{Imaging constraints and state reconstruction}
\label{ssec:limits}

Classical phase-retrieval methods combine magnitude data with constraints
\citep{gerchberg_saxton1972,fienup1982,marchesini2007,shechtman2015phasereview}; ptychography
adds measurement diversity through overlapping illuminations
\citep{rodenburg2004,maiden2009epie}. These examples motivate the question
``which acquisition makes the desired property identifiable?'' They do
not furnish interchangeable algorithms for arbitrary quantum states.
An explicit algebraic bridge is low-rank matrix recovery. For a complex
field $f$ and known measurement vectors $a_j$, PhaseLift
\citep{candes2013phaselift} uses measured intensities $b_j$ as constraints
on a lifted matrix:
\begin{equation}
 \begin{aligned}
 b_j&=|a_j^\dagger f|^2=\operatorname{Tr}(a_ja_j^\dagger F),\\
 F&=ff^\dagger\succeq0,
 \end{aligned}
 \label{eq:phaselift}
\end{equation}
Here $F=ff^\dagger$ is the field's outer-product matrix,
$\operatorname{Tr}$ sums diagonal entries, and $F\succeq0$ means that
$F$ is positive semidefinite. The quadratic intensity constraints are
linear in this rank-one matrix. Convex relaxation recovers $F$ under
suitable measurement conditions, fixing $f$ up to global phase.
Quantum tomography likewise infers a positive matrix $\rho$ of trace
one from probabilities $p_j=\operatorname{Tr}(M_j\rho)$.
The positive matrices $M_j$, with $0\preceq M_j\preceq I$, specify the
measured events; $p_j$ is the probability of event $j$.
Section~\ref{ssec:mixed} develops this density-matrix description.
Compressed-sensing quantum tomography exploits low rank
\citep{gross2010compressed,flammia2012compressed}. The shared structure
is inference over positive matrices. Recovery depends on the measurement
ensemble; estimating quantum probabilities also prices copies and
finite-shot uncertainty. Table~\ref{tab:measurement_bounds} summarizes
the operational correspondences.
The companion's \href{https://quantum-works.elettra.eu/seeing-amplitudes/routes/identifiability/}{measurement route}
makes these connections interactive, from phase and shot-budget controls
to the one-qubit geometry of positive-matrix constraints.

\begin{table*}[!htbp]
\centering\small
\caption{Useful imaging correspondences and their operational boundaries.}
\label{tab:measurement_bounds}
\begin{tabular}{@{}>{\raggedright\arraybackslash}p{3.1cm}
                  >{\raggedright\arraybackslash}p{4.5cm}
                  >{\raggedright\arraybackslash}p{5.6cm}@{}}
\toprule
\textbf{Imaging concept} & \textbf{What transfers} & \textbf{Boundary}\\
\midrule
Intensity detection & Squared magnitudes omit phase relationships & A projective readout generally changes the prepared quantum copy.\\
Acquisition diversity & Different settings reveal different properties & Statistics of the original state in incompatible settings require separate preparations; arbitrary unknown states cannot be cloned \citep{wootters1982}.\\
Reference interference & Recombination makes an overlap observable & Ancillary readout presumes controlled operations and is not unrestricted access to a classical field.\\
Counting statistics & Independent samples support uncertainty estimates & A detector frame integrates many events; one register shot supplies one basis index.\\
Low-rank recovery & Lifted intensities and tomography give linear constraints on a positive matrix & Measurement ensembles, normalization and sample costs determine the recovery guarantee.\\
\bottomrule
\end{tabular}
\end{table*}

%% file: figures/phase_settings.tex
\begin{compactfigure}
\centering
\begin{adjustbox}{max width=\linewidth}
\begin{tikzpicture}
 \foreach \y/\name/\fn in {0/X/\cos,-1.35/Y/\sin}{
   \node[anchor=south west,font=\figtitle,inner sep=0pt] at (-0.85,\y+0.42) {$\name$ setting};
   \node[anchor=east] at (0,\y) {$\ket{\psi_\phi}$};
   \draw[qwire] (0,\y) -- (3.1,\y);
   \draw[cwire] (3.4,\y) -- (3.85,\y);
   \node[gate] at (2.0,\y) {$H$};
   \pic at (3.1,\y) {meter};
   \node[anchor=west] at (3.9,\y) {$2p_{\name}(0)-1=\fn\phi$};
 }
 \node[gate] at (0.85,-1.35) {$S^{\dagger}$};
\end{tikzpicture}
\end{adjustbox}
\caption{Analysis circuits for two independently acquired ensembles.
The $Y$ setting applies $S^{\dagger}$ before $H$, so its matrix is
$HS^{\dagger}$. Every shot starts with a fresh source preparation;
the two statistics are combined classically to infer the phase.}
\label{fig:phase_diversity}
\end{compactfigure}

%% file: figures/phase_efficiency.tex
\begin{compactfigure}
\centering
\begin{adjustbox}{max width=\linewidth}
\begin{tikzpicture}
\begin{axis}[width=\columnwidth,height=4.1cm,
 xmin=0,xmax=2,ymin=0.7,ymax=2.12,
 xlabel={Phase $\phi$},ylabel={Variance ratio $R$},
 xtick={0,.5,1,1.5,2},
 xticklabels={$0$,$\pi/2$,$\pi$,$3\pi/2$,$2\pi$},
 axis lines=left,axis line style={FigureMuted,-},tick style={FigureMuted},
 ymajorgrids,grid style={FigureMuted!15,line width=.3pt},
 ytick={1,1.5,2},clip=false]
\addplot[FigureMuted,line width=.6pt,dash pattern=on 3pt off 2pt,domain=0:2] {1};
\addplot[FigureTeal,line width=1.1pt] table[x=phi_over_pi,y=variance_ratio_XY,col sep=comma]
 {results/phase_efficiency.csv};
\addplot[only marks,mark=*,mark size=1.7pt,FigureOchre,mark options={draw=white,line width=.4pt}]
 coordinates {(\PhaseExampleOverPi,\PhaseXYVarianceRatioPlot)};
\node[font=\figsmall,anchor=north,text=FigureOchre!85!black,inner sep=1.5pt]
 at (axis cs:\PhaseExampleOverPi,0.985) {$\phi=\PhaseTruth$};
\node[font=\figsmall,anchor=north east,text=FigureMuted,inner sep=1.5pt]
 at (axis cs:2,0.985) {local $1/(2N)$ benchmark};
\end{axis}
\end{tikzpicture}
\end{adjustbox}
\caption{Delta-method phase variance divided by the local $1/(2N)$
benchmark for $N$ X and $N$ Y measurements. The curve extends the
regular-phase expression continuously to $\phi=k\pi/2$, where
deterministic outcomes require separate regularity treatment.
The marker locates $\phi=\PhaseTruth$ rad.}
\label{fig:phase_efficiency}
\end{compactfigure}

%% file: sections/05_composites.tex
\section{Composite systems and classical simulation}
\label{sec:entanglement}

This section develops composite-state structure through coefficient matrices and their factorizations.

\subsection{A coefficient matrix and its singular values}
\label{ssec:coeff}

Represent a two-qubit pure state by its amplitude array and reshape it
into a coefficient matrix:
\begin{equation}
 \begin{aligned}
 \ket\psi&=\sum_{i,j=0}^{1}c_{ij}\ket{i}\ket{j},\\
 C&=\begin{pmatrix}c_{00}&c_{01}\\c_{10}&c_{11}\end{pmatrix},\qquad
 \|C\|_F^2=1.
 \end{aligned}
\end{equation}
The row index labels the first qubit and the column index the second:
$c_{01}$, for example, is the amplitude of $\ket{01}$.
The Frobenius normalization $\|C\|_F^2=\sum_{i,j}|c_{ij}|^2=1$ is the
state-vector normalization, unchanged by reshaping.
For local amplitude columns $\vec a$ and $\vec b$, the tensor product
in Eq.~\eqref{eq:kron} gives $C=\vec a\vec b^{\top}$, a rank-one outer
product. Conversely, rank one
allows that factorization; rank greater than one means the joint pure
state is \emph{entangled}. For this two-qubit matrix,
$\det C=c_{00}c_{11}-c_{01}c_{10}$ is zero for a product state and
nonzero for an entangled state. The imaging correspondence is the
failure of a separable kernel representation,
now with an assigned physical tensor-product partition. A nonseparable
classical image is not thereby quantum-entangled: the matrix argument
applies to coefficients of a physical joint quantum state, not to every
array on which an SVD can be performed.

In the SVD $C=L\Sigma R^{\dagger}$, $L$ and $R$ are unitary and
$\Sigma$ is diagonal, with nonnegative singular values $\sigma_r$.
Expanding the matrix entries gives the Schmidt
decomposition
\begin{equation}
 \ket\psi=\sum_r\sigma_r\ket{\ell_r}\otimes\ket{r_r^*},
 \qquad \sum_r\sigma_r^2=1.
 \label{eq:schmidt}
\end{equation}
Here $\ell_r$ and $r_r$ are columns of $L$ and $R$; the conjugation of
the second vector follows from $R^{\dagger}$ in the complex SVD.
Thus the Schmidt rank is the coefficient matrix rank. For a pure
bipartite state it diagnoses entanglement, while the spectrum quantifies
it, for example through
\begin{equation}
 S=-\sum_r\sigma_r^2\log_2\sigma_r^2.
 \label{eq:schmidt_entropy}
\end{equation}
These statements refer to the specified bipartition. They are invariant
under separate unitary basis changes on its two subsystems, but changing
which degrees of freedom constitute the subsystems changes the question.
Nor is matrix rank a general test of entanglement for mixed states.

\subsection{The Bell state and an indistinguishable first record}

The Bell state
\begin{equation}
 \ket{\Phi^+}=\frac{\ket{00}+\ket{11}}{\sqrt2},\qquad
 C_{\Phi^+}=\frac{I}{\sqrt2}
\end{equation}
has Schmidt rank two, equal singular values $1/\sqrt2$, and entropy one
bit. Its preparation uses a Hadamard on the first wire followed by the
controlled-\textnormal{NOT} introduced in Section~\ref{sec:transform}:
\begin{equation}
 \ket{00}\xrightarrow{H\otimes I}
 \frac{\ket{00}+\ket{10}}{\sqrt2}
 \xrightarrow{\mathrm{CNOT}}\ket{\Phi^+}.
 \label{eq:bell}
\end{equation}
The comparison state $\ket{++}$ has all four amplitudes equal to $1/2$,
but its coefficient matrix has rank one. Equal participation of many
basis states therefore does not itself diagnose entanglement.
Figure~\ref{fig:product_vs_bell} pairs the Bell preparation circuit with
the two readout settings used below.

Computational-basis measurement of $\ket{\Phi^+}$ yields $00$ or $11$,
each with probability $1/2$. A classical random preparation of
$\ket{00}$ or $\ket{11}$ produces exactly the same record. Measuring
both wires in $X$, by placing $H$ before each readout, separates them:
the Bell preparation again gives only $00$ and $11$, whereas the random
preparation gives all four outcomes equally. Correlation in one basis
does not expose all of the joint state's structure.

\input{figures/bell_statistics.tex}

\subsection{Mixed states and a minimal noise model}
\label{ssec:mixed}

Density matrices extend the amplitude description to classical mixtures
and subsystems. A pure state has
$\rho=\ket\psi\bra\psi$; a preparation ensemble has
$\rho=\sum_j w_j\ket{\psi_j}\bra{\psi_j}$, with $w_j\ge0$ and
$\sum_jw_j=1$. The Bell state and the preceding random preparation are
\begin{align}
 \rho_{\Phi^+}&=\tfrac12\big(\ket{00}\bra{00}+\ket{11}\bra{11}
                     +\ket{00}\bra{11}+\ket{11}\bra{00}\big),
 \notag\\
 \rho_{\mathrm{mix}}&=\tfrac12\big(\ket{00}\bra{00}
                                 +\ket{11}\bra{11}\big).
 \label{eq:bell_mixture}
\end{align}
Their computational-basis diagonals agree; the off-diagonal terms carry
the coherence revealed by $XX$. Tracing out the second qubit sums over
its unobserved basis index:
\[
 \rho_A=\sum_{b=0}^{1}(I\otimes\bra b)\rho_{\Phi^+}
                         (I\otimes\ket b)=I/2.
\]
The other qubit has the same reduced state by symmetry. Each is mixed,
even though the joint state is pure and no laboratory noise is present.

For density matrices, unitary evolution is $\rho\mapsto U\rho U^{\dagger}$
and outcome probabilities are
$p(k)=\operatorname{Tr}(\Pi_k\rho)$ for measurement projectors $\Pi_k$.
More general noise processes are channels
$\rho\mapsto\sum_\mu K_\mu\rho K_\mu^{\dagger}$ with
$\sum_\mu K_\mu^{\dagger}K_\mu=I$ \citep{nielsen_chuang}. One useful
illustration is the dephasing channel
\begin{equation}
 \mathcal D_\eta(\rho)=\frac{1+\eta}{2}\rho
                 +\frac{1-\eta}{2}Z\rho Z,\qquad 0\le\eta\le1.
 \label{eq:dephasing}
\end{equation}
It preserves diagonal entries and multiplies off-diagonal entries by
$\eta$. Applied to $\ket{\psi_\phi}$ before the analysis Hadamard, it
changes Eq.~\eqref{eq:p1} to
\begin{equation}
 \Pr(1)=\frac{1-\eta\cos\phi}{2}.
 \label{eq:dephased_fringe}
\end{equation}
The imaging interpretation is a reduction in interference contrast.
Complete dephasing ($\eta=0$) removes the phase information from that
fringe; increasing the number of shots cannot restore absent contrast.
Likewise, dephasing one Bell qubit interpolates between
$\rho_{\Phi^+}$ and $\rho_{\mathrm{mix}}$, with
$\langle X\otimes X\rangle=\eta$.

This is a stated illustrative channel, not a fitted device model.
Real hardware also has relaxation, leakage, readout error, crosstalk and
drift, and a published gate-error number does not determine $\eta$.

\subsection{Compressibility and the cost of classical simulation}
\label{sec:exponential}

Reshaping an $n$-qubit vector across a chosen left/right partition gives
the same SVD construction at larger scale. Small Schmidt ranks across
the contiguous cuts of a chosen qubit ordering permit a compact matrix
product state representation \citep{orus2014,vidal2003simulation}.
The bond dimension is the size of the internal index linking neighboring
tensors; it bounds the Schmidt rank retained across that cut.
With $\sigma_1\ge\sigma_2\ge\cdots$, the Eckart--Young theorem \citep{eckart1936approximation} identifies a best rank-$r$ approximation
$C_r$ in Frobenius norm, with
\begin{equation}
 \min_{\operatorname{rank}(D)\le r}\|C-D\|_F^2
       =\|C-C_r\|_F^2=\sum_{j>r}\sigma_j^2.
 \label{eq:eckart_young}
\end{equation}
For the unnormalized truncated state, this is the squared discarded
vector norm. Renormalizing the retained
state changes the error measure; it does not change the discarded weight. Approximate simulation depends on this spectrum
and its accumulated error, not only on the number of nonzero values.
For local circuits whose intermediate bond dimensions remain bounded
polynomially, tensor-network updates can be efficient. A low-rank final
state alone does not guarantee that every intermediate computation was
easy.

Entanglement growth can obstruct this representation without proving
general classical intractability. Stabilizer circuits, built from $H$,
$S=P(\pi/2)$ and controlled-\textnormal{NOT}, on computational-basis inputs
with Pauli measurements, admit efficient classical simulation despite
potentially extensive entanglement \citep{aaronson2004stabilizer}.
Their simulator tracks an algebraic description rather than an amplitude
array. Adding arbitrary phase rotations removes this particular
guarantee, but does not make every resulting circuit hard. The gate set,
structure, observable and requested accuracy all matter.

A dense state-vector simulator nevertheless has an unmistakable memory
cost. At $16$ bytes per double-precision complex amplitude it uses
$16\,2^n$ bytes before working storage. For $n=20,30,40,50$, these are
$16$ MiB, $16$ GiB, $16$ TiB and $16$ PiB, respectively. Those figures
price one representation, not every classical simulation method.
Some methods trade time for memory or exploit structure without
materializing the vector. A short quantum circuit is itself a compact
classical specification; calculating its amplitudes or sampling its
outputs can be much harder than storing that specification.

The computational question is whether useful output statistics of an
efficiently implementable quantum process admit efficient classical
computation under the same access assumptions. Establishing a separation
requires task-specific analysis and classical baselines. Both worked
examples here are small enough for exact classical simulation.

%% file: figures/bell_statistics.tex
\begin{figure*}[!tp]
\centering
\begin{adjustbox}{max width=\linewidth}
\begin{tikzpicture}
 \draw[qwire] (0,-0.75) -- (3.0,-0.75);
 \draw[qwire] (0,-1.8) -- (3.0,-1.8);
 \node[anchor=east] at (0,-0.75) {$\ket0$};
 \node[anchor=east] at (0,-1.8) {$\ket0$};
 \node[gate] at (0.85,-0.75) {$H$};
 \draw[qwire] (2.05,-0.75) -- (2.05,-1.8);
 \node[ctrl] at (2.05,-0.75) {};
 \pic at (2.05,-1.8) {targ};
 \node at (1.5,-2.85) {$\ket{\Phi^+}=\dfrac{\ket{00}+\ket{11}}{\sqrt2}$};
 \foreach \col/\state in {0/Bell state,1/Classical mixture}{
  \node[font=\figtitle] at ({8.1+5.0*\col},0.95) {\state};}
 \foreach \row/\set/\pa/\pb/\pc/\pd/\qa/\qb/\qc/\qd in
   {0/ZZ/0.5/0/0/0.5/0.5/0/0/0.5, 1/XX/0.5/0/0/0.5/0.25/0.25/0.25/0.25}{
  \node[anchor=east,align=right] at (6.05,{-0.3-2.35*\row}) {$\set$\\readout};
  \foreach \col/\a/\b/\c/\d in {0/\pa/\pb/\pc/\pd,1/\qa/\qb/\qc/\qd}{
   \begin{scope}[shift={({6.3+5.0*\col},{-0.95-2.35*\row})}]
    \draw[FigureMuted!60,line width=.45pt] (0,0) -- (3.6,0);
    \foreach \k/\p/\lab in {0/\a/00,1/\b/01,2/\c/10,3/\d/11}{
     \pgfmathsetmacro{\barcol}{(\row==1 && \col==1) ? 1 : 0}
     \ifdim\barcol pt>0.5pt\def\bc{FigureOchre!85}\else\def\bc{FigureTeal!80}\fi
     \fill[\bc] ({0.22+0.85*\k},0) rectangle ++(0.62,{2.5*\p});
     \node[font=\figsmall,anchor=north] at ({0.53+0.85*\k},-0.02) {\lab};
     \node[font=\figsmall,anchor=south,text=FigureMuted] at ({0.53+0.85*\k},{2.5*\p})
      {\ifdim\p pt=0pt $0$\else\ifdim\p pt=0.25pt $1/4$\else$1/2$\fi\fi};}
   \end{scope}}}
\end{tikzpicture}
\end{adjustbox}
\caption{Bell preparation and exact ideal outcome probabilities.
$ZZ$ denotes computational readout of both qubits, and $XX$ denotes an
$H$ before each readout. The second setting distinguishes coherence
that the first record does not reveal (ochre).}
\label{fig:product_vs_bell}
\end{figure*}
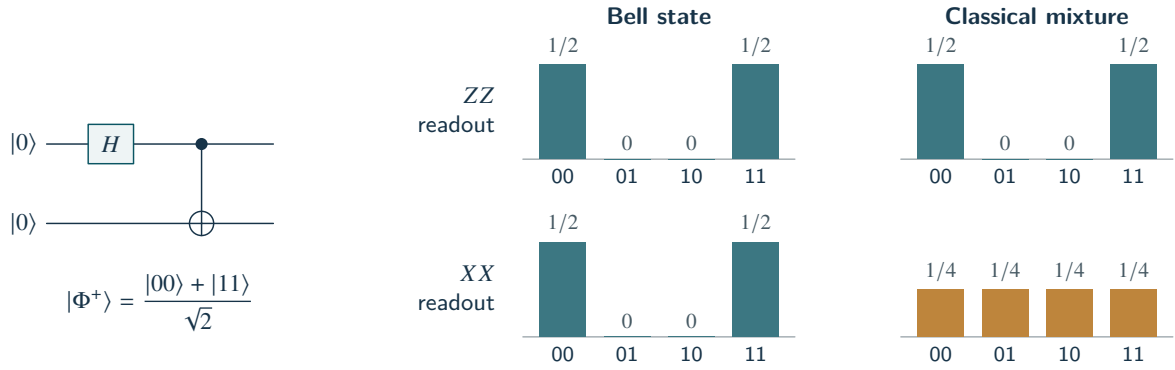

%% file: sections/07_imaging_example.tex
\section{From an imaging objective to a sampled decision}
\label{sec:outlook}

The preceding sections describe how quantum computation represents and
transforms amplitudes. An imaging problem starts with observed data and
an image model, from which we specify an objective over unknown labels
and a decision to be returned.
This section carries a small binary segmentation through the complete
chain
\[
\begin{gathered}
\text{data and model}\;\rightarrow\;\text{binary objective}
\;\rightarrow\;\text{cost operator}\\
\rightarrow\;\text{circuit}\;\rightarrow\;\text{decoded samples}.
\end{gathered}
\]
The small chain exposes the complete representation and its interpretation,
with the classical baseline visible throughout.

\subsection{From pixel labels to a quadratic objective}
\label{ssec:bridge2}

Let $x_i\in\{0,1\}$ denote the unknown label of pixel $i$, as in the
decision-variable encoding of Section~\ref{ssec:encoding}.
We minimize an objective $E$ combining data-fit costs $d_i(x_i)$ with
spatial regularization over neighboring pixel pairs $\mathcal N$:
\begin{equation}
 E(\vec{x})=\sum_i d_i(x_i)
       +\sum_{(i,j)\in\mathcal N}w_{ij}[x_i\ne x_j],
 \qquad w_{ij}\ge0.
 \label{eq:seg}
\end{equation}
The weight $w_{ij}$ sets the cost of disagreeing labels on pair $(i,j)$;
the bracket $[x_i\ne x_j]$ is one for disagreement and zero otherwise
(equal to binary XOR, $x_i\oplus x_j$).
Each undirected pair is included once. Since $x_i^2=x_i$ and
$[x_i\ne x_j]=x_i+x_j-2x_ix_j$, this is a quadratic unconstrained binary
optimization (QUBO) problem,
\begin{equation}
 \begin{aligned}
 E(\vec{x})&=C+\sum_i a_i x_i+\sum_{i<j}b_{ij}x_ix_j
            =C+\vec{x}^{\mathsf T}Q\vec{x},\\
 Q_{ii}&=a_i,\qquad Q_{ij}=b_{ij}\ (i<j),\qquad Q_{ji}=0\ (i<j).
 \end{aligned}
 \label{eq:qubo}
\end{equation}
The upper-triangular convention counts every pair once. A symmetric
matrix can represent the same objective, but its off-diagonal entries
must then be $b_{ij}/2$. Mixing these conventions changes the problem
\citep{lucas2014ising,glover2019qubo}.

For the nonnegative disagreement costs in Eq.~\eqref{eq:seg}, the
two-label pairwise objective is submodular: agreement has no greater
pair cost than disagreement. An $s$--$t$ minimum cut solves
it exactly in polynomial time, with the data costs attached to the
terminals and the spatial penalties to the graph edges
\citep{boykov2001,kolmogorov2004}. This remains the relevant classical
reference after a quantum formulation is introduced. QUBO expresses a
common representation; it does not erase the exploitable structure of a
particular instance.

For a concrete one-dimensional example, take four synthetic normalized
pixel values on a chain:
\begin{equation}
 \begin{aligned}
 \vec{g}&=(0.9,\,0.8,\,0.45,\,0.85),\\
 d_i(x_i)&=(g_i-x_i)^2,\qquad w_{12}=w_{23}=w_{34}=0.2.
 \end{aligned}
 \label{eq:toydata}
\end{equation}
Thresholding at $0.5$ returns $1101$,
with energy $\QThresholdEnergy$. Enumeration of all sixteen assignments
gives the unique objective minimizer $1111$, with energy $\QOptEnergy$.
Figure~\ref{fig:four_pixel_model} shows how the spatial penalty changes
the third label and maps each labeling to a basis state. The change is
a consequence of the stated objective.

\input{figures/four_pixel_model.tex}

Expanding the terms gives
\begin{equation}
 Q=\begin{pmatrix}
 -0.6&-0.4&0&0\\
 0&-0.2&-0.4&0\\
 0&0&0.5&-0.4\\
 0&0&0&-0.5
 \end{pmatrix},\quad
 \begin{aligned}E(\vec{x})&=2.375\\&\quad+\vec{x}^{\mathsf T}Q\vec{x}.\end{aligned}
 \label{eq:qubo4}
\end{equation}
For example, the unary coefficient at pixel 2 is
$(1-2g_2)+0.2+0.2=-0.2$, while edge $(2,3)$ contributes
$-0.4x_2x_3$. The supplementary reference checks these coefficients against
the original objective for all sixteen assignments and independently
solves the chain by dynamic programming.

\subsection{From an objective to a cost circuit}
\label{ssec:cost_circuit}

Here the encoding is the \emph{decision-variable encoding} of
Section~\ref{sec:state}: four labels use four qubits. The basis state
$\ket{x_1x_2x_3x_4}$ denotes one segmentation, and the amplitudes define
a search distribution over sixteen segmentations. These amplitudes do
not encode the four pixel intensities. We use qubit 1 as the most
significant bit, so $1101$ has array index 13.

Replace each binary variable $x_i$ in Eq.~\eqref{eq:qubo} by the
binary-label projector $n_i=(I-Z_i)/2$, which has eigenvalue $x_i$ on
$\ket{x_i}$. This substitution turns the classical objective into a
cost operator, or \emph{cost Hamiltonian}, expressed in Pauli terms:
\begin{equation}
 H_E=c_0I+\sum_i h_i Z_i+\sum_{i<j}J_{ij}Z_iZ_j,
 \qquad H_E\ket{\vec{x}}=E(\vec{x})\ket{\vec{x}}.
 \label{eq:ising}
\end{equation}
\begin{equation}
 \begin{aligned}
 c_0&=C+\tfrac12\sum_i a_i+\tfrac14\sum_{i<j}b_{ij},\\
 h_i&=-\tfrac12a_i-\tfrac14\sum_{j\ne i}b_{\min(i,j),\max(i,j)},
 \qquad J_{ij}=\tfrac14b_{ij}.
 \end{aligned}
 \label{eq:qubo_pauli_general}
\end{equation}
Thus the concrete cost operator is
\begin{equation}
 \begin{aligned}
 H_E&=1.675I+0.4Z_1+0.3Z_2-0.05Z_3+0.35Z_4\\
 &\quad-0.1(Z_1Z_2+Z_2Z_3+Z_3Z_4).
 \end{aligned}
 \label{eq:pauliH}
\end{equation}
Its diagonal is the complete energy table. This is the Ising form of
the binary objective, using spins with values $\pm1$. Our convention
uses additive field and coupling coefficients as written in $H_E$; an
overall minus sign on fields and couplings requires the corresponding
sign conversion.

With the objective fixed, we now choose a parameterized sampling circuit.
A variational method updates its parameters classically to reduce a
measured objective. We use a standard one-layer quantum approximate
optimization algorithm (QAOA) construction with an $X$ mixer
\citep{farhi2014qaoa,cerezo2021vqa}. The adjustable parameters $\gamma$
and $\beta$ set the cost and mixer evolutions:
\begin{equation}
 \begin{aligned}
 \ket{\psi(\gamma,\beta)}&=
 e^{-\ii\beta\sum_iX_i}e^{-\ii\gamma H_E}\ket{+}^{\otimes4},\\
 p_{\gamma,\beta}(\vec{x})&=
 |\langle\vec{x}|\psi(\gamma,\beta)\rangle|^2.
 \end{aligned}
 \label{eq:toyqaoa}
\end{equation}
The initial $\ket{+}^{\otimes4}$ is a product of four $\ket{+}$ states,
with equal amplitudes over all sixteen labelings. The cost layer assigns
phase $-\gamma E(\vec{x})$ to each amplitude, leaving the uniform measurement
probabilities unchanged until the mixer creates interference.
Because the cost terms commute, this layer
factorizes into four single-qubit $Z$ evolutions and three two-qubit
$ZZ$ evolutions, rather than sixteen independently specified basis-state
phase operations. The mixer factorizes into four $X$ evolutions. This
structure connects the spatial graph directly to a local circuit
\citep{farhi2014qaoa}.

Using the single-qubit rotations of Eq.~\eqref{eq:rotations}, define
\begin{equation}
 R_{ZZ}(\theta)=e^{-\ii\theta Z\otimes Z/2}.
 \label{eq:rotation_convention}
\end{equation}
The cost angles are therefore $2\gamma h_i$ and $2\gamma J_{ij}$, and
the mixer angle is $2\beta$. The factor of two is an implementation
convention with mathematical consequences. For example,
$R_{ZZ}(\theta)$ decomposes as CNOT--$R_Z(\theta)$--CNOT, with the
$R_Z$ on the target. The constant $c_0$ contributes only the global
phase $e^{-\ii\gamma c_0}$ and can be omitted from this circuit while
retaining it in reported energies. Figure~\ref{fig:toy_circuit} shows
the resulting preparation, cost, mixer and readout stages.

\input{figures/qaoa_layer.tex}

\subsection{From a sampling law to an imaging decision}
\label{ssec:decision_sampling}

At fixed angles, a computational-basis shot returns a labeling
$\vec{x}$. The value $z_i=1-2x_i\in\{-1,+1\}$ is its $Z_i$ eigenvalue, and
$z_iz_j=(-1)^{x_i+x_j}$ gives the $Z_iZ_j$ eigenvalue. Every term in
$H_E$ is diagonal in this same basis, so a single acquisition setting
estimates all fields and couplings. Including the retained constant
$c_0$, their weighted sum on each shot is exactly the original energy
$E(\vec{x})$. For $N$ independent, identically
prepared shots, the finite-shot energy estimator is therefore
\begin{equation}
 \widehat{\langle H_E\rangle}
       =\frac{1}{N}\sum_{r=1}^{N}E(\vec{x}^{(r)}).
 \label{eq:sampled_cost}
\end{equation}
For samples from the fixed ideal distribution $p_{\gamma,\beta}$,
the estimator is unbiased for $\langle H_E\rangle$. Its variance is
the single-shot energy variance divided by $N$, including correlations
between terms obtained from the same shot. Selecting angles from noisy
estimates introduces a separate parameter-selection problem.

For the reproducible reference, evaluate the exact state-vector
expectation on a fixed $241\times121$ grid of cost angles
$\gamma\in[0,2\pi]$ and mixer angles $\beta\in[0,\pi]$, respectively,
and retain its best point. The grid dimensions are numerical design choices.
The selected point is $(\gamma,\beta)=(\QGamma,\QBeta)$. The
$[0,2\pi]$ cost-angle window keeps the illustrative grid compact while
allowing the leading coefficients to generate appreciable relative
phases. It is a restricted search, not a continuous-global-optimum
certificate. For the current
energies, whose differences are integer multiples of $0.1$ with greatest
common divisor $0.1$, the cost evolution has period $20\pi$ modulo
global phase. The accompanying reference also checks a full-period grid
at the same spacing and obtains the same minimum grid-evaluated expected energy.
Rescaling
$H_E$ also rescales the corresponding cost-angle domain.

\begin{table*}[!htbp]
\centering\small
\caption{Distinct summaries of ideal depth-1 QAOA at the selected grid
point $(\gamma,\beta)=(\QGamma,\QBeta)$. Probabilities and
expectations are exact state-vector quantities; finite-shot observations
fluctuate around them.}
\label{tab:toy_results}
\begin{tabular}{lll}
\toprule
Quantity & Uniform preparation & One QAOA layer\\
\midrule
Expected energy & $\QUniformEnergy$ & $\QMeanEnergy$\\
$\Pr(1111)$, the objective optimum & $0.0625$ & $\QOptProb$\\
$\Pr(1101)$, the threshold labeling & $0.0625$ & $\QRunnerProb$\\
Shots for $\ge99\%$ chance of at least one optimum & $\UniformShots$ & $\QShots$\\
\bottomrule
\end{tabular}
\end{table*}

\input{figures/qaoa_distribution.tex}

Table~\ref{tab:toy_results} summarizes the ideal calculation, and
Figure~\ref{fig:toy_distribution} displays its full sampling law.
The unary data costs favour the threshold labeling $1101$, whereas the
boundary penalty makes $1111$ the full-objective optimum. The one-layer
mixer coherently recombines cost-phased amplitudes rather than ranking
assignments by energy; at the selected angles that interference gives
$1101$ more weight than $1111$, since minimizing the mean does not
require the optimum to be modal.

Three outputs should be distinguished. First, the \emph{expected energy}
decreases from $\QUniformEnergy$ to $\QMeanEnergy$, but remains above
the deterministic threshold labeling's $\QThresholdEnergy$. Second,
the \emph{mode} is $1101$, not the objective optimum. Third, a
\emph{best-of-$m$ decoder} evaluates the original objective on the
sampled labels and retains the lowest-energy one. If each independently
prepared shot has optimum probability $p_\star$, then
\begin{equation}
 P_m=1-(1-p_\star)^m.
 \label{eq:bestofm}
\end{equation}
Here $P_m$ is the probability of observing an optimum at least once,
and $p_\star\simeq\QOptProb$, so $m=\QShots$ reaches at least
$99\%$, compared with $\UniformShots$ for uniform sampling. These
success probabilities concern the known objective optimum. A
computational comparison must also count state preparation, circuit
execution, parameter search and classical cost evaluations against the
exact classical solver already available for this instance.

During the parameter search, the classical optimizer changes the angles;
during decoding, it interprets samples at fixed angles. The reference
uses exact simulated expectations for its search rather than
Eq.~\eqref{eq:sampled_cost}; hardware training would additionally involve
sampling and execution errors. Keeping these
stages separate explains how a mathematically valid quantum formulation
can coexist with an imperfect algorithmic decision, even before device
noise is considered. That is the practical lesson the imaging formulation
adds to the circuit description. Alternatives such as conditional
value-at-risk (CVaR) training emphasize a low-energy tail rather than the
mean \citep{cvar2020}; the angles reported here remain selected by the
mean energy.
The companion's \href{https://quantum-works.elettra.eu/seeing-amplitudes/routes/decision/\#x-seg-decode}{decision panel}
compares these summaries on the same sampling law; its
\href{https://quantum-works.elettra.eu/seeing-amplitudes/routes/decision/\#x-hardware}{hardware panel}
separately displays the archived six-pixel local event.

\input{sections/06b_hardware_case.tex}

%% file: figures/four_pixel_model.tex
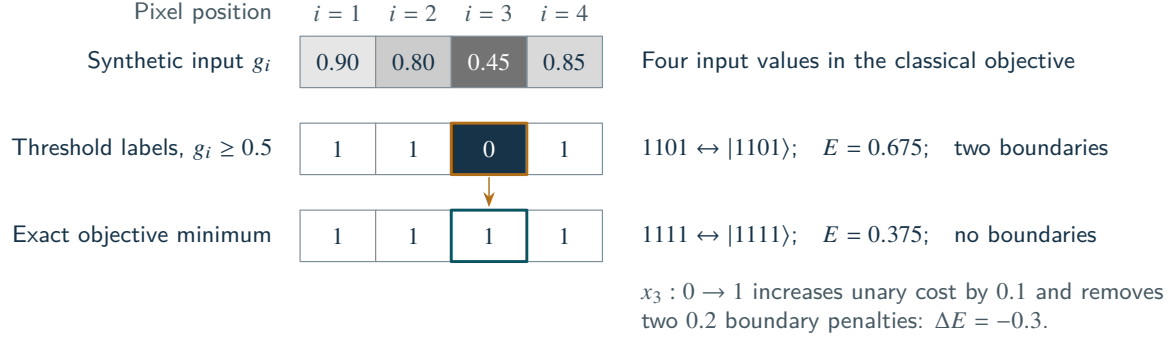
\begin{figure*}[!tp]
\centering
\begin{adjustbox}{max width=\linewidth}
\begin{tikzpicture}[x=1cm,y=1cm,
 cell/.style={draw=FigureInk!55,line width=.45pt,minimum width=1.0cm,minimum height=0.68cm,inner sep=0pt},
 rowlabel/.style={anchor=east,align=right},
 result/.style={anchor=west,align=left}]
\node[rowlabel,text=FigureMuted] at (-0.25,3.1) {Pixel position};
\foreach \i in {1,2,3,4}{\node[text=FigureMuted] at (\i-0.5,3.1) {$i=\i$};}
\node[rowlabel] at (-0.25,2.45) {Synthetic input $g_i$};
\foreach \xx/\shade/\val in {0.5/10/0.90,1.5/20/0.80,3.5/15/0.85}{\node[cell,fill=black!\shade] at (\xx,2.45) {$\val$};}
\node[cell,fill=black!55,text=white] at (2.5,2.45) {$0.45$};
\node[result] at (4.4,2.45) {Four input values in the classical objective};
\node[rowlabel] at (-0.25,1.3) {Threshold labels, $g_i\ge0.5$};
\foreach \xx in {0.5,1.5,3.5}{\node[cell,fill=white] at (\xx,1.3) {$1$};}
\node[cell,fill=FigureInk,text=white] at (2.5,1.3) {$0$};
\draw[FigureOchre,line width=1.1pt] (2.0,0.96) rectangle (3.0,1.64);
\node[result] at (4.4,1.3) {$1101\leftrightarrow\ket{1101}$;\quad $E=\QThresholdEnergy$;\quad two boundaries};
\node[rowlabel] at (-0.25,0.15) {Exact objective minimum};
\foreach \xx in {0.5,1.5,2.5,3.5}{\node[cell,fill=white] at (\xx,0.15) {$1$};}
\draw[FigureTeal,line width=1.1pt] (2.0,-0.19) rectangle (3.0,0.49);
\draw[fig arrow,draw=FigureOchre] (2.5,0.92) -- (2.5,0.53);
\node[result] at (4.4,0.15) {$1111\leftrightarrow\ket{1111}$;\quad $E=\QOptEnergy$;\quad no boundaries};
\node[anchor=north west,align=left,text=FigureMuted] at (4.4,-0.35)
 {$x_3:0\to1$ increases unary cost by $0.1$ and removes\\two $0.2$ boundary penalties: $\Delta E=-0.3$.};
\end{tikzpicture}
\end{adjustbox}
\caption{The four-pixel input and two classical labeling decisions.
The top strip represents $g_i$ by grayscale, with brighter cells denoting
larger values. The outlined third label exposes the tradeoff between
unary fit and boundary cost. A labeling specifies a four-qubit basis
state; $1111$ is the exact classical reference.}
\label{fig:four_pixel_model}
\end{figure*}

%% file: figures/qaoa_layer.tex
\begin{figure*}[!tp]
\centering
\begin{adjustbox}{max width=\linewidth}
\begin{tikzpicture}[
 zz/.style={draw=FigureTeal,line width=.55pt,fill=FigureSoft}]
\def\dy{0.85}
\foreach \xa/\xb/\t/\c in {0.5/1.45/prepare/FigureMuted!7,1.6/10.35/cost phases/FigureTeal!7,
  10.55/12.55/mix/FigureOchre!9,12.75/14.0/sample/FigureMuted!7}{
 \fill[\c,rounded corners=2pt] (\xa,0.5) rectangle (\xb,{-3*\dy-0.45});
 \node[anchor=south,font=\figsmall,text=FigureMuted] at ({(\xa+\xb)/2},0.5) {\t};}
\foreach \q in {1,2,3,4}{
 \pgfmathsetmacro{\yy}{-(\q-1)*\dy}
 \draw[qwire] (0,\yy) -- (13.4,\yy);
 \draw[cwire] (13.7,\yy) -- (14.3,\yy);
 \node[anchor=east] at (0,\yy) {$\ket0_{\q}$};
 \node[gate] at (0.97,\yy) {$H$};
 \node[gate] at (2.65,\yy) {$R_Z(2\gamma h_{\q})$};
 \node[gate] at (11.55,\yy) {$R_X(2\beta)$};
 \pic at (13.4,\yy) {meter};}
\foreach \a/\b/\xx in {1/2/5.0,2/3/7.12,3/4/9.24}{
 \pgfmathsetmacro{\ya}{-(\a-1)*\dy}\pgfmathsetmacro{\yb}{-(\b-1)*\dy}
 \draw[zz] (\xx-0.98,\ya+0.28) rectangle (\xx+0.98,\yb-0.28);
 \node at (\xx,{(\ya+\yb)/2}) {$R_{ZZ}(2\gamma J_{\a\b})$};}
\end{tikzpicture}
\end{adjustbox}
\caption{One layer for the four-pixel chain. Each box spanning two wires
is one $R_{ZZ}$ gate with the angle shown; the final symbols denote
computational-basis measurement, and double lines carry the classical outcomes.}
\label{fig:toy_circuit}
\end{figure*}
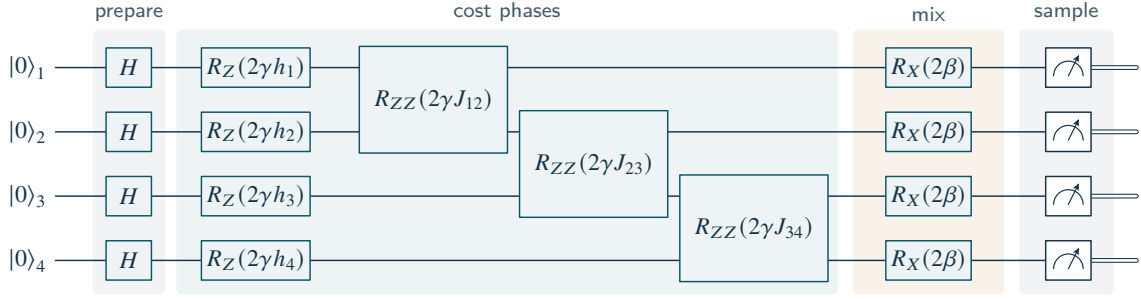

%% file: figures/qaoa_distribution.tex
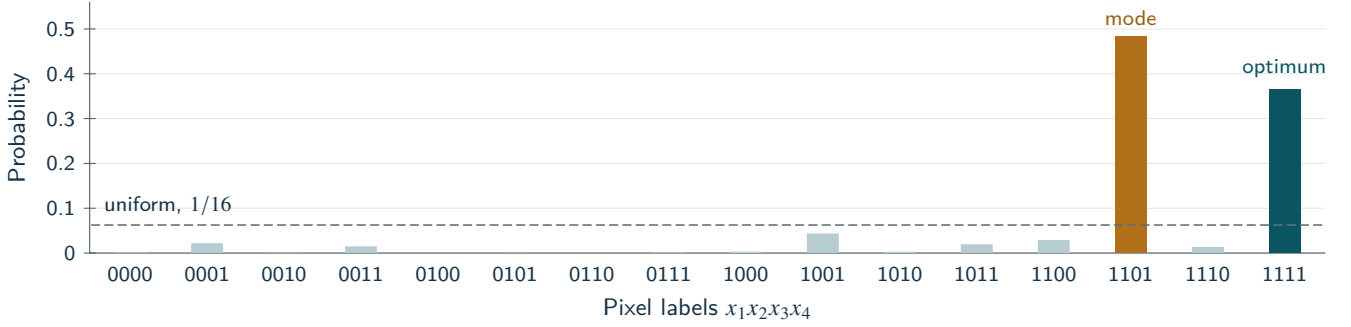
\begin{figure*}[!tp]
\centering
\begin{adjustbox}{max width=\linewidth}
\begin{tikzpicture}
\begin{axis}[width=\linewidth,height=4.9cm,ybar,bar width=12pt,bar shift=0pt,
 ymin=0,ymax=0.56,ylabel={Probability},xlabel={Pixel labels $x_1x_2x_3x_4$},
 xtick={0,...,15},xticklabels={0000,0001,0010,0011,0100,0101,0110,0111,
 1000,1001,1010,1011,1100,1101,1110,1111},xtick style={draw=none},
 axis lines=left,axis line style={FigureMuted,-},tick style={FigureMuted},
 ymajorgrids,grid style={FigureMuted!15,line width=.3pt},
 ytick={0,0.1,0.2,0.3,0.4,0.5},enlarge x limits=0.035,clip=false]
 \addplot[fill=FigureTeal!30,draw=none] table[x=index,y=probability,col sep=comma,
  restrict expr to domain={\thisrow{index}}{0:12}] {results/qaoa_distribution.csv};
 \addplot[fill=FigureTeal!30,draw=none] table[x=index,y=probability,col sep=comma,
  restrict expr to domain={\thisrow{index}}{14:14}] {results/qaoa_distribution.csv};
 \addplot[fill=FigureOchre,draw=none] table[x=index,y=probability,col sep=comma,
  restrict expr to domain={\thisrow{index}}{13:13}] {results/qaoa_distribution.csv};
 \addplot[fill=FigureTeal,draw=none] table[x=index,y=probability,col sep=comma,
  restrict expr to domain={\thisrow{index}}{15:15}] {results/qaoa_distribution.csv};
 \draw[FigureInk!70,line width=.6pt,dash pattern=on 3pt off 2pt] (axis cs:-0.5,0.0625) -- (axis cs:15.5,0.0625);
 \node[font=\figsmall,anchor=south west,text=FigureInk] at (axis cs:-0.45,0.066) {uniform, $1/16$};
 \node[font=\figsmall,anchor=south,text=FigureOchre!85!black] at (axis cs:13,0.49) {mode};
 \node[font=\figsmall,anchor=south,text=FigureTeal] at (axis cs:15,0.372) {optimum};
\end{axis}
\end{tikzpicture}
\end{adjustbox}
\caption{Ideal depth-1 QAOA distribution at the selected grid point
$(\gamma,\beta)=(\QGamma,\QBeta)$. The
dashed line is uniform sampling. The mode $1101$ has greater probability
than the unique objective optimum $1111$.}
\label{fig:toy_distribution}
\end{figure*}

%% file: sections/06b_hardware_case.tex
\subsection{Physical execution and the local boundary decision}
\label{ssec:hardware_case}
A physical realization translates the logical task into native operations
and records the mappings needed to interpret the returned bits.
A related provenance study provides archived measurements from a matched
campaign on IBM Kingston and IQM Garnet
\citep[Supplementary Section~S5]{KourousiasProvenance2026}.
This is a separate six-variable conditional computed-tomography (CT) task
with synthetic input intensities, rather than a hardware execution of the
four-pixel example above. Its objective, scale, angles and decoder were
fixed across two layouts and three acquisition blocks on each processor.
Table~\ref{tab:archivedhardware} contains all twelve matched contexts.

Six active pixels form a path. Conditioning on the exterior labels gives
\begin{align}
 E_U(x)&=C+\sum_{i\in U}a_i^{\mathrm{eff}}x_i
       +\lambda\sum_{\substack{(i,j)\in\mathcal N\\i,j\in U}}
                       (x_i\oplus x_j),\notag\\
 a_i^{\mathrm{eff}}&=-(I_i-t)
       +\lambda\sum_{\substack{j\notin U\\(i,j)\in\mathcal N}}
                    (1-2\bar x_j).
 \label{eq:ct_conditional}
\end{align}
Here $I_i$ is the normalized intensity, $t$ the threshold,
$\lambda\ge0$ the smoothness weight and $\bar x_j$ an exterior label.
For binary labels, $x_i\oplus x_j=[x_i\ne x_j]$ is the same
disagreement indicator as in Eq.~\eqref{eq:seg}. Every fixed-boundary interaction is retained,
and each undirected edge is counted once \citep{kourousias2026jinst}.
The numerical case uses $t=0.5$ and $\lambda=0.06$. Its exact coefficients,
Hamiltonian scale, frozen angles, native circuits and row-to-count links
are specified by \nolinkurl{results/hardware/ct6_hardware.json} and
\nolinkurl{hardware_specification.md} in this article's supporting
package. That compact extract links each row to the original campaign
records \citep{QuantumProvenanceDataset}.

The decoder answers a local boundary question. The two central active
pixels straddle the designed synthetic interface, with intensities on
opposite sides of the threshold. Fixed before acquisition,
$d(x)=(x_3,x_4)$ asks whether that edge is labeled background-to-foreground;
its reference output is $01$. The other four variables still participate
in the objective and evolution, while the reported law sums over them:
\begin{equation}
 q(01)=\sum_{x:d(x)=01}p(x)\ge p(000111).
 \label{eq:decoder_event}
\end{equation}
This event contains sixteen assignments, including the unique full
minimizer $000111$. Uniform bit sampling gives $1/64$ for that minimizer
and $1/4$ for the local event. The two columns therefore concern
success at different levels of detail.

The ideal probabilities are $\HardwareIdealOptimum$ and
$\HardwareIdealDecoded$, respectively. Across the measured contexts,
the full-assignment frequency ranges from $\HardwareOptimumMin$ to
$\HardwareOptimumMax$, and the local-event frequency from
$\HardwareDecodedMin$ to $\HardwareDecodedMax$. The local label $01$
is the empirical mode in every context. These observations distinguish
a local edge decision from complete patch refinement; their usefulness
depends on which output the application requires.
On Kingston, layout~0 gives full-minimizer frequencies
$\HardwareIBMZeroOptMin$--$\HardwareIBMZeroOptMax$, compared with
$\HardwareIBMOneOptMin$--$\HardwareIBMOneOptMax$ for layout~1.
The layout with higher full-minimizer frequencies has lower local-event
frequencies, while $01$ remains modal. This descriptive contrast neither
isolates a layout effect nor establishes equivalent decoded laws.

\begin{compacttable}
\caption{CT boundary measurements from the matched campaign reported in
\citep[Supplementary Section~S5]{KourousiasProvenance2026} and archived in
\citep{QuantumProvenanceDataset}. All twelve contexts are shown separately.
$n$ is the number of shots; $\hat p$ and $\hat q$ are empirical
frequencies of the full assignment and sixteen-assignment local event.
A layout specifies a physical-qubit mapping; a block is a separate
acquisition. Their labels are provider-specific. Supplementary Archive~S1
gives pointwise 95\% Wilson intervals under independent, identically
distributed shots within each context, without assuming independence
between contexts or simultaneous coverage.}
\label{tab:archivedhardware}
\centering\small
\begin{tabular}{@{}lccrrr@{}}
\toprule
Device & Layout & Block & $n$ & $\hat p(000111)$ & $\hat q(01)$\\
\midrule
\multicolumn{4}{l}{Uniform bit reference} & $1/64$ & $1/4$\\
\multicolumn{4}{l}{Ideal logical reference} & \HardwareIdealOptimumTable & \HardwareIdealDecodedTable\\
\midrule
\input results/hardware_context_rows.tex
\bottomrule
\end{tabular}
\end{compacttable}

%% file: results/hardware_context_rows.tex
IBM & 0 & 0 & 4,096 & 0.3608 & 0.7729 \\
IBM & 0 & 1 & 4,096 & 0.3457 & 0.7693 \\
IBM & 0 & 2 & 4,096 & 0.3582 & 0.7805 \\
IBM & 1 & 0 & 4,096 & 0.6106 & 0.7571 \\
IBM & 1 & 1 & 4,096 & 0.6094 & 0.7517 \\
IBM & 1 & 2 & 4,096 & 0.6072 & 0.7412 \\
IQM & 0 & 0 & 2,048 & 0.5454 & 0.7090 \\
IQM & 0 & 1 & 2,048 & 0.5249 & 0.7012 \\
IQM & 0 & 2 & 2,048 & 0.4941 & 0.6948 \\
IQM & 1 & 0 & 2,048 & 0.5806 & 0.7598 \\
IQM & 1 & 1 & 2,048 & 0.5532 & 0.7578 \\
IQM & 1 & 2 & 2,048 & 0.5522 & 0.7476 \\

%% file: sections/06_algorithms.tex
\section{Algorithmic consequences of the access model}
\label{sec:algorithms}
Unknown-phase inference (Sections~\ref{ssec:bridge1}--\ref{ssec:phase_information})
and four-pixel segmentation (Section~\ref{sec:outlook}) make different
outputs accessible: a phase estimate and a sampled labeling, respectively.
Broader algorithms combine the same preparation,
transformation and inference operations with additional access assumptions.
Three patterns locate the preceding derivations within that wider setting
\citep{nielsen_chuang,montanaro2016,auyeung2024}.

\subsection{Phase marking and amplitude amplification}
For candidate index $k$, a binary test $f(k)\in\{0,1\}$ and an auxiliary
bit $a$, a reversible predicate oracle acts as
$U_f\ket{k}\ket{a}=\ket{k}\ket{a\oplus f(k)}$, where $\oplus$ is addition
modulo two. Preparing the auxiliary qubit in $\ket{-}$ gives
\begin{equation}
 U_f\ket{k}\ket{-}=(-1)^{f(k)}\ket{k}\ket{-}.
 \label{eq:phase_kickback}
\end{equation}
This \emph{phase kickback} changes signs without changing immediate
computational-basis probabilities. Across a superposition of candidates,
the signs are relative phases, as in the controlled-branch discussion of
Section~\ref{sec:transform}. Grover search alternates such marking
with reflection about the uniform state \citep{grover1996}. For one marked
item among $N$ candidates, $O(\sqrt N)$ oracle calls give constant success
probability, followed by measurement and classical verification. The
comparison with $O(N)$ independent uniform trials prices oracle work
inside each execution, not merely the number of returned samples.
An imaging predicate such as ``the residual is below a threshold'' must
itself have a reversible implementation; evaluating a forward model
inside that predicate can determine the cost. This is the same
phase-to-output distinction exposed by the cost layer and mixer in
Section~\ref{sec:outlook}.

\subsection{Coherent phase accumulation and Fourier readout}
Quantum phase estimation extends the ancilla interference of
Section~\ref{ssec:recovery}. For $U\ket{u}=e^{2\pi\ii\theta}\ket{u}$,
prepare $m$ ancillas and controlled powers of $U$. With $R=2^m$, their
state before inverse Fourier readout is
\begin{equation}
 \frac{1}{\sqrt R}\sum_{k=0}^{R-1}e^{2\pi\ii k\theta}\ket{k}\ket{u}.
 \label{eq:qpe_phase_state}
\end{equation}
The inverse quantum Fourier transform yields
\begin{equation}
 \Pr(y\mid\theta)=\left|\frac{1}{R}\sum_{k=0}^{R-1}
 e^{2\pi\ii k(\theta-y/R)}\right|^2,
 \label{eq:qpe_distribution}
\end{equation}
which peaks near $y/R=\theta$ \citep{cleve1998}. An exact Fourier circuit
uses $O(m^2)$ elementary gates, but controlled $U^{2^j}$ generally costs
$2^j$ uses of $U$ unless additional structure is available. Precision is
paid for by coherent evolution as well as readout. This differs from
repeating the fixed single-phase experiment of
Section~\ref{ssec:bridge1}; its $N^{-1/2}$ empirical scaling is not a
bound on every quantum estimation protocol.

A classical fast Fourier transform returns a readable coefficient array;
the quantum transform produces a state used by later operations or a
specific measurement. Comparing their transform counts without matching
these outputs compares different tasks. Shor's order-finding algorithm
uses efficiently implementable modular arithmetic and classical
postprocessing to turn this interference into a useful result
\citep{shor1997}.

\subsection{A solution state and an imaging functional}
For invertible Hermitian $A=\sum_j\lambda_j\ket{u_j}\bra{u_j}$ and
$\ket{b}=\sum_j b_j\ket{u_j}$, a quantum linear-system method targets
\begin{equation}
 \ket{x}=\frac{\sum_j b_j\lambda_j^{-1}\ket{u_j}}
 {\left\|\sum_j b_j\lambda_j^{-1}\ket{u_j}\right\|}.
 \label{eq:linear_solution_state}
\end{equation}
The original Harrow--Hassidim--Lloyd construction uses eigenvalue
estimation, a controlled flag rotation and uncomputation, conditioned on
success \citep{hhl2009}. It does not apply an arbitrary nonunitary inverse
as a deterministic gate. A non-Hermitian problem may be embedded in a
larger Hermitian one, with access and conditioning costs retained.

For imaging, the important distinction is between the normalized state
$\ket{x}$ and a classical reconstruction $\vec{x}$. Preparing
$\ket b$, accessing $A$, conditioning, precision and success probability
all contribute; obtaining the solution norm or its entries is a further
output task \citep{aaronson2015fineprint,auyeung2024}. A selected overlap
or functional may be the appropriate scientific output instead.
Modern block encodings place a scaled matrix inside an ancilla-selected
block of a unitary, and quantum singular value transformation constructs
polynomial functions of its singular values \citep{gilyen2019qsvt,martyn2021qsvt}.
This connects spectral filtering to the SVD perspective, provided the
block construction and requested output have suitable costs.

Access assumptions also matter on the classical side. Tang's
quantum-inspired recommendation algorithm illustrates how strong
sampling and query access can enable classical low-rank methods
\citep{tang2019}. It is not a general dequantization of imaging inverse
problems. It motivates specifying the same available data structures,
preprocessing and output requirement before asserting a separation.

The variational circuit in Section~\ref{sec:outlook} belongs to another
regime: a classical optimizer updates parameters using sampled costs
\citep{cerezo2021vqa}. Ideal adiabatic computation instead changes a
Hamiltonian continuously under gap- and schedule-dependent conditions
\citep{albash2018}; physical annealing also includes environmental and
thermal effects \citep{kadowaki1998,hauke2020}. An Ising objective specifies
a target for these methods, not an efficient route to its minimizer.

%% file: sections/08_practical.tex
\section{Resource account, implications and outlook}
\label{sec:practical}
An imaging representation becomes a computational proposal only when its
access assumptions and returned quantity are specified. The preceding
examples make those choices concrete: a phase, a full labeling and a
pair of selected labels are different outputs, even when related by
classical processing.

\subsection{Matching the scientific task and the resource account}
The end-to-end viewpoint of Dalzell \emph{et al.}\ combines primitive
operations into the cost of the requested output \citep{dalzell2025endtoend}.
Table~\ref{tab:resource_account} translates that viewpoint into imaging
specifications. Data access itself is an implementation choice: quantum
random-access memory (QRAM) must be priced with its controls and the
classical resources that could support a competing computation
\citep{jaques2025qram}.
\begin{table*}[!htbp]
\centering\small
\caption{An input-to-output account for a quantum imaging proposal.
The same output quality and access assumptions must be used for the
classical comparison.}
\label{tab:resource_account}
\begin{tabular}{@{}>{\raggedright\arraybackslash}p{0.19\textwidth}
 >{\raggedright\arraybackslash}p{0.75\textwidth}@{}}
\toprule
Stage & Required specification\\
\midrule
Representation & Meaning of a basis index and an amplitude; retained scale,
metadata and model assumptions; the estimate or decision required.\\[3pt]
Preparation and access & State preparation and repetition; matrix, oracle,
controlled operation or inverse access; construction and preprocessing costs.\\[3pt]
Transformation & Local and entangling gates, ancillas, routing, precision
and permitted approximation; efficient synthesis of the stated operator.\\[3pt]
Measurement & Observable or distribution; settings and preparations;
uncertainty, drift assumptions and classical decoding.\\[3pt]
Classical computation & Parameter search, forward-model evaluations,
verification and decoding; equal accounting for preprocessing.\\[3pt]
Memory and data movement & Representation stored or streamed on each route;
working memory, data access and reuse; no assumption that a classical method
materializes every array or dense operator.\\[3pt]
Baseline and validity & Exploitable classical structure; output quality;
independent truth where physical accuracy is claimed; scope of the comparison.\\
\bottomrule
\end{tabular}
\end{table*}
The classical reference also changes what a quantum result means. The
four-pixel objective is exactly solvable by a minimum cut; its quantum
encoding demonstrates a representation and sampling mechanism rather
than a reason to replace that solver. Conditional boundary refinement
preserves the same distinction. Agreement with a model minimizer is
separate from agreement with independently established physical truth.

Streaming is another access model. Quantum oracle sketching processes
classical samples incrementally to support approximate coherent queries
\citep{zhao2026oraclesketching}; practical analyses examine its implementation
costs \citep{markidis2026oraclesketching}. Compact-memory comparisons must
therefore specify sample consumption, updates and coherent execution,
alongside the classical streaming baseline.

Characterization separates preparation, gate and readout contributions
\citep{hashim2025benchmarking}. Fault-tolerant resource estimates additionally
price error correction and physical spacetime costs; Gidney and Eker\aa's
factoring study provides a concrete example \citep{gidney2021factoring}.
These costs extend the logical accounting used here.

\subsection{Outlook and continuation}
A productive research direction is to design the computation around the
imaging quantity actually required. A full reconstructed image, a chosen
correlation, an uncertainty-relevant statistic and a model-selection
decision impose different readout requirements. Avoiding full-array
recovery can alter the opportunity, but only when the reduced output
serves the same scientific question on both computational routes.
The CT example in Section~\ref{ssec:hardware_case} makes this concrete:
a local boundary question and complete patch refinement concern different
events in the same measurement record. Choosing the output is therefore
part of the scientific specification before it becomes a readout or
resource-budget decision.

Operator-based algorithms connect spectral filtering and regularization
to potential fault-tolerant inverse-problem subroutines
\citep{martyn2021qsvt}. Near-term variational methods permit small,
explicitly testable discrete formulations while exposing the interaction
between training, execution and decoding \citep{cerezo2021vqa}.
Computational imaging contributes forward models, structured classical
solvers, acquisition design and realistic quality criteria to both.
Useful benchmarks preserve these application choices while separating
an exact or well-characterized classical reference, the ideal quantum
algorithm and physical observations.

For further study, standard texts develop general measurements, channels
and quantum information \citep{nielsen_chuang,mermin_qcs,kitaev_shen_vyalyi}.
Control and microwave accounts connect these to robustness and devices
\citep{berberich2024control,bardin2020microwave}; the scientific-computing
survey develops algorithmic opportunities \citep{auyeung2024}.
The supplied implementation provides an independently checkable reference
before translation into another software stack.

\section{Conclusions}
\label{sec:summary}
The encoding comparison distinguishes input values stored in amplitudes
from amplitudes assigned to complete segmentations. Computational-basis
readout returns a location in the first case and a labeling in the second.
For the unknown source phase in
Sections~\ref{ssec:bridge1}--\ref{ssec:phase_information}, acquisition
diversity becomes a choice of complementary readout settings. The
Fisher-information account separates ambiguity resolution, local precision
and expenditure on diagnostic copies. For a joint pure state, SVD turns
coefficient-matrix factorization into an entanglement criterion across a
specified partition; efficient simulation additionally depends on circuit
structure and intermediate states. The four-pixel example then carries
data-fit and spatial-regularization costs into a circuit and three
distinct outputs: expected energy, modal labeling and best sampled
labeling. Archived six-pixel measurements extend this output distinction
to a local boundary event, with execution contexts separated and
pointwise frequency intervals supplied in Supplementary Archive~S1.

The examples establish representation and inference correspondences,
rather than quantum advantage. The four-pixel parameter search is
restricted, and its objective has an exact classical solver; the
six-pixel hardware observations illustrate one fixed task rather than
device rankings or equivalent decoded distributions. Practical inverse
problems additionally require justified data-access assumptions, full
resource accounting and validation against physical truth.

%% file: main.bbl
\begin{thebibliography}{10}
\providecommand{\url}[1]{#1}
\csname url@samestyle\endcsname
\providecommand{\newblock}{\relax}
\providecommand{\bibinfo}[2]{#2}
\providecommand{\BIBentrySTDinterwordspacing}{\spaceskip=0pt\relax}
\providecommand{\BIBentryALTinterwordstretchfactor}{4}
\providecommand{\BIBentryALTinterwordspacing}{\spaceskip=\fontdimen2\font plus
\BIBentryALTinterwordstretchfactor\fontdimen3\font minus
  \fontdimen4\font\relax}
\providecommand{\BIBforeignlanguage}[2]{{%
\expandafter\ifx\csname l@#1\endcsname\relax
\typeout{** WARNING: IEEEtran.bst: No hyphenation pattern has been}%
\typeout{** loaded for the language `#1'. Using the pattern for}%
\typeout{** the default language instead.}%
\else
\language=\csname l@#1\endcsname
\fi
#2}}
\providecommand{\BIBdecl}{\relax}
\BIBdecl

\bibitem{nielsen_chuang}
M.~A. Nielsen and I.~L. Chuang, \emph{Quantum Computation and Quantum
  Information}.\hskip 1em plus 0.5em minus 0.4em\relax Cambridge: Cambridge
  University Press, 2010, 10th anniversary edition.

\bibitem{mermin_qcs}
N.~D. Mermin, \emph{Quantum Computer Science: An Introduction}.\hskip 1em plus
  0.5em minus 0.4em\relax Cambridge: Cambridge University Press, 2007.

\bibitem{kitaev_shen_vyalyi}
A.~Y. Kitaev, A.~H. Shen, and M.~N. Vyalyi, \emph{Classical and Quantum
  Computation}, ser. Graduate Studies in Mathematics.\hskip 1em plus 0.5em
  minus 0.4em\relax Providence, RI: American Mathematical Society, 2002,
  vol.~47.

\bibitem{bardin2020microwave}
J.~C. Bardin, D.~Sank, O.~Naaman, and E.~Jeffrey, ``Quantum computing: An
  introduction for microwave engineers,'' \emph{IEEE Microwave Magazine},
  vol.~21, no.~8, pp. 24--44, 2020, doi:
  \href{https://doi.org/10.1109/MMM.2020.2993475}{\nolinkurl{10.1109/MMM.2020.2993475}}.

\bibitem{berberich2024control}
J.~Berberich and D.~Fink, ``Quantum computing through the lens of control: A
  tutorial introduction,'' \emph{IEEE Control Systems}, vol.~44, no.~6, pp.
  24--49, 2024, doi:
  \href{https://doi.org/10.1109/MCS.2024.3466448}{\nolinkurl{10.1109/MCS.2024.3466448}}.

\bibitem{auyeung2024}
R.~Au-Yeung, B.~Camino, O.~Rathore, and V.~Kendon, ``Quantum algorithms for
  scientific computing,'' \emph{Reports on Progress in Physics}, vol.~87,
  no.~11, p. 116001, 2024, doi:
  \href{https://doi.org/10.1088/1361-6633/ad85f0}{\nolinkurl{10.1088/1361-6633/ad85f0}}.

\bibitem{beardow2026metaphors}
C.~Beardow and P.~J. Stappers, ``Towards metaphors for quantum computational
  thinking,'' \emph{EPJ Quantum Technology}, vol.~13, p.~55, 2026, doi:
  \href{https://doi.org/10.1140/epjqt/s40507-026-00502-1}{\nolinkurl{10.1140/epjqt/s40507-026-00502-1}}.

\bibitem{jun2025segmentation}
K.~Jun and H.~Lee, ``Quantum optimization algorithms for {CT} image
  segmentation from {X}-ray data,'' \emph{Scientific Reports}, vol.~15, p.
  20649, 2025, doi:
  \href{https://doi.org/10.1038/s41598-025-08453-w}{\nolinkurl{10.1038/s41598-025-08453-w}}.

\bibitem{kourousias2026xrf}
G.~Kourousias, S.~Carrato, F.~Guzzi, A.~Gianoncelli, A.~Hafner, R.~Pugliese,
  and P.~Dimitrakis, ``Quantum-native formulations for {X}-ray fluorescence:
  spectral overlap, sparse elemental identification and spatial inference,''
  \emph{X-Ray Spectrometry}, 2026, in press. doi:
  \href{https://doi.org/10.1002/xrs.70141}{\nolinkurl{10.1002/xrs.70141}}.

\bibitem{le2011frqi}
P.~Q. Le, F.~Dong, and K.~Hirota, ``A flexible representation of quantum images
  for polynomial preparation, image compression, and processing operations,''
  \emph{Quantum Information Processing}, vol.~10, pp. 63--84, 2011, doi:
  \href{https://doi.org/10.1007/s11128-010-0177-y}{\nolinkurl{10.1007/s11128-010-0177-y}}.

\bibitem{zhang2013neqr}
Y.~Zhang, K.~Lu, Y.~Gao, and M.~Wang, ``{NEQR}: a novel enhanced quantum
  representation of digital images,'' \emph{Quantum Information Processing},
  vol.~12, pp. 2833--2860, 2013, doi:
  \href{https://doi.org/10.1007/s11128-013-0567-z}{\nolinkurl{10.1007/s11128-013-0567-z}}.

\bibitem{pittman1995ghost}
T.~B. Pittman, Y.~H. Shih, D.~V. Strekalov, and A.~V. Sergienko, ``Optical
  imaging by means of two-photon quantum entanglement,'' \emph{Physical Review
  A}, vol.~52, p. R3429, 1995, doi:
  \href{https://doi.org/10.1103/PhysRevA.52.R3429}{\nolinkurl{10.1103/PhysRevA.52.R3429}}.

\bibitem{bennink2002classicalghost}
R.~S. Bennink, S.~J. Bentley, and R.~W. Boyd, ``\,`{Two-Photon}' coincidence
  imaging with a classical source,'' \emph{Physical Review Letters}, vol.~89,
  p. 113601, 2002, doi:
  \href{https://doi.org/10.1103/PhysRevLett.89.113601}{\nolinkurl{10.1103/PhysRevLett.89.113601}}.

\bibitem{brida2010imaging}
G.~Brida, M.~Genovese, and I.~Ruo~Berchera, ``Experimental realization of
  sub-shot-noise quantum imaging,'' \emph{Nature Photonics}, vol.~4, pp.
  227--230, 2010, doi:
  \href{https://doi.org/10.1038/nphoton.2010.29}{\nolinkurl{10.1038/nphoton.2010.29}}.

\bibitem{KourousiasProvenance2026}
\BIBentryALTinterwordspacing
G.~Kourousias, S.~Carrato, R.~Pugliese, A.~Hafner, F.~Guzzi, M.~Bill{\`e},
  F.~Bill{\`e}, and P.~Dimitrakis, ``Sufficient quantum provenance: retained
  fields and certified recording precision,'' arXiv:2610.09931, 2026, doi:
  \href{https://doi.org/10.48550/arXiv.2610.09931}{\nolinkurl{10.48550/arXiv.2610.09931}}.
  [Online]. Available: \url{https://arxiv.org/abs/2610.09931}
\BIBentrySTDinterwordspacing

\bibitem{seeingamplitudes2026}
\BIBentryALTinterwordspacing
G.~Kourousias, S.~Carrato, R.~Pugliese, P.~Dimitrakis, and A.~Carini, ``{Seeing
  Amplitudes}: Representations, measurement and inference,'' Interactive
  browser-based laboratory, 2026. [Online]. Available:
  \url{https://quantum-works.elettra.eu/seeing-amplitudes/}
\BIBentrySTDinterwordspacing

\bibitem{plesch2011}
M.~Plesch and {\v{C}}.~Brukner, ``Quantum-state preparation with universal gate
  decompositions,'' \emph{Physical Review A}, vol.~83, no.~3, p. 032302, 2011,
  doi:
  \href{https://doi.org/10.1103/PhysRevA.83.032302}{\nolinkurl{10.1103/PhysRevA.83.032302}}.

\bibitem{goodman2005}
J.~W. Goodman, \emph{Introduction to Fourier Optics}, 3rd~ed.\hskip 1em plus
  0.5em minus 0.4em\relax Englewood, CO: Roberts and Company, 2005.

\bibitem{huang2020shadows}
H.-Y. Huang, R.~Kueng, and J.~Preskill, ``Predicting many properties of a
  quantum system from very few measurements,'' \emph{Nature Physics}, vol.~16,
  no.~10, pp. 1050--1057, 2020, doi:
  \href{https://doi.org/10.1038/s41567-020-0932-7}{\nolinkurl{10.1038/s41567-020-0932-7}}.

\bibitem{kay1993estimation}
S.~M. Kay, \emph{Fundamentals of Statistical Signal Processing, Vol. {I}:
  Estimation Theory}.\hskip 1em plus 0.5em minus 0.4em\relax Englewood Cliffs,
  NJ: Prentice Hall, 1993.

\bibitem{braunstein1994geometry}
S.~L. Braunstein and C.~M. Caves, ``Statistical distance and the geometry of
  quantum states,'' \emph{Physical Review Letters}, vol.~72, no.~22, pp.
  3439--3443, 1994, doi:
  \href{https://doi.org/10.1103/PhysRevLett.72.3439}{\nolinkurl{10.1103/PhysRevLett.72.3439}}.

\bibitem{gabor1948}
D.~Gabor, ``A new microscopic principle,'' \emph{Nature}, vol. 161, pp.
  777--778, 1948, doi:
  \href{https://doi.org/10.1038/161777a0}{\nolinkurl{10.1038/161777a0}}.

\bibitem{gerchberg_saxton1972}
R.~W. Gerchberg and W.~O. Saxton, ``A practical algorithm for the determination
  of phase from image and diffraction plane pictures,'' \emph{Optik}, vol.~35,
  no.~2, pp. 237--246, 1972.

\bibitem{fienup1982}
J.~R. Fienup, ``Phase retrieval algorithms: A comparison,'' \emph{Applied
  Optics}, vol.~21, no.~15, pp. 2758--2769, 1982, doi:
  \href{https://doi.org/10.1364/AO.21.002758}{\nolinkurl{10.1364/AO.21.002758}}.

\bibitem{marchesini2007}
S.~Marchesini, ``Invited article: A unified evaluation of iterative projection
  algorithms for phase retrieval,'' \emph{Review of Scientific Instruments},
  vol.~78, no.~1, p. 011301, 2007, doi:
  \href{https://doi.org/10.1063/1.2403783}{\nolinkurl{10.1063/1.2403783}}.

\bibitem{shechtman2015phasereview}
Y.~Shechtman, Y.~C. Eldar, O.~Cohen, H.~N. Chapman, J.~Miao, and M.~Segev,
  ``Phase retrieval with application to optical imaging: A contemporary
  overview,'' \emph{IEEE Signal Processing Magazine}, vol.~32, no.~3, pp.
  87--109, 2015, doi:
  \href{https://doi.org/10.1109/MSP.2014.2352673}{\nolinkurl{10.1109/MSP.2014.2352673}}.

\bibitem{rodenburg2004}
J.~M. Rodenburg and H.~M.~L. Faulkner, ``A phase retrieval algorithm for
  shifting illumination,'' \emph{Applied Physics Letters}, vol.~85, no.~20, pp.
  4795--4797, 2004, doi:
  \href{https://doi.org/10.1063/1.1823034}{\nolinkurl{10.1063/1.1823034}}.

\bibitem{maiden2009epie}
A.~M. Maiden and J.~M. Rodenburg, ``An improved ptychographical phase retrieval
  algorithm for diffractive imaging,'' \emph{Ultramicroscopy}, vol. 109,
  no.~10, pp. 1256--1262, 2009, doi:
  \href{https://doi.org/10.1016/j.ultramic.2009.05.012}{\nolinkurl{10.1016/j.ultramic.2009.05.012}}.

\bibitem{candes2013phaselift}
E.~J. Cand{\`e}s, T.~Strohmer, and V.~Voroninski, ``{PhaseLift}: Exact and
  stable signal recovery from magnitude measurements via convex programming,''
  \emph{Communications on Pure and Applied Mathematics}, vol.~66, no.~8, pp.
  1241--1274, 2013, doi:
  \href{https://doi.org/10.1002/cpa.21432}{\nolinkurl{10.1002/cpa.21432}}.

\bibitem{gross2010compressed}
D.~Gross, Y.-K. Liu, S.~T. Flammia, S.~Becker, and J.~Eisert, ``Quantum state
  tomography via compressed sensing,'' \emph{Physical Review Letters}, vol.
  105, no.~15, p. 150401, 2010, doi:
  \href{https://doi.org/10.1103/PhysRevLett.105.150401}{\nolinkurl{10.1103/PhysRevLett.105.150401}}.

\bibitem{flammia2012compressed}
S.~T. Flammia, D.~Gross, Y.-K. Liu, and J.~Eisert, ``Quantum tomography via
  compressed sensing: Error bounds, sample complexity, and efficient
  estimators,'' \emph{New Journal of Physics}, vol.~14, no.~9, p. 095022, 2012,
  doi:
  \href{https://doi.org/10.1088/1367-2630/14/9/095022}{\nolinkurl{10.1088/1367-2630/14/9/095022}}.

\bibitem{wootters1982}
W.~K. Wootters and W.~H. Zurek, ``A single quantum cannot be cloned,''
  \emph{Nature}, vol. 299, pp. 802--803, 1982, doi:
  \href{https://doi.org/10.1038/299802a0}{\nolinkurl{10.1038/299802a0}}.

\bibitem{orus2014}
R.~Or{\'u}s, ``A practical introduction to tensor networks: Matrix product
  states and projected entangled pair states,'' \emph{Annals of Physics}, vol.
  349, pp. 117--158, 2014, doi:
  \href{https://doi.org/10.1016/j.aop.2014.06.013}{\nolinkurl{10.1016/j.aop.2014.06.013}}.

\bibitem{vidal2003simulation}
G.~Vidal, ``Efficient classical simulation of slightly entangled quantum
  computations,'' \emph{Physical Review Letters}, vol.~91, no.~14, p. 147902,
  2003, doi:
  \href{https://doi.org/10.1103/PhysRevLett.91.147902}{\nolinkurl{10.1103/PhysRevLett.91.147902}}.

\bibitem{eckart1936approximation}
C.~Eckart and G.~Young, ``The approximation of one matrix by another of lower
  rank,'' \emph{Psychometrika}, vol.~1, no.~3, pp. 211--218, 1936, doi:
  \href{https://doi.org/10.1007/BF02288367}{\nolinkurl{10.1007/BF02288367}}.

\bibitem{aaronson2004stabilizer}
S.~Aaronson and D.~Gottesman, ``Improved simulation of stabilizer circuits,''
  \emph{Physical Review A}, vol.~70, no.~5, p. 052328, 2004, doi:
  \href{https://doi.org/10.1103/PhysRevA.70.052328}{\nolinkurl{10.1103/PhysRevA.70.052328}}.

\bibitem{lucas2014ising}
A.~Lucas, ``Ising formulations of many {NP} problems,'' \emph{Frontiers in
  Physics}, vol.~2, p.~5, 2014, doi:
  \href{https://doi.org/10.3389/fphy.2014.00005}{\nolinkurl{10.3389/fphy.2014.00005}}.

\bibitem{glover2019qubo}
F.~Glover, G.~Kochenberger, and Y.~Du, ``Quantum bridge analytics {I}: A
  tutorial on formulating and using {QUBO} models,'' \emph{4OR -- A Quarterly
  Journal of Operations Research}, vol.~17, no.~4, pp. 335--371, 2019, doi:
  \href{https://doi.org/10.1007/s10288-019-00424-y}{\nolinkurl{10.1007/s10288-019-00424-y}}.

\bibitem{boykov2001}
Y.~Boykov, O.~Veksler, and R.~Zabih, ``Fast approximate energy minimization via
  graph cuts,'' \emph{IEEE Transactions on Pattern Analysis and Machine
  Intelligence}, vol.~23, no.~11, pp. 1222--1239, 2001, doi:
  \href{https://doi.org/10.1109/34.969114}{\nolinkurl{10.1109/34.969114}}.

\bibitem{kolmogorov2004}
V.~Kolmogorov and R.~Zabih, ``What energy functions can be minimized via graph
  cuts?'' \emph{IEEE Transactions on Pattern Analysis and Machine
  Intelligence}, vol.~26, no.~2, pp. 147--159, 2004, doi:
  \href{https://doi.org/10.1109/TPAMI.2004.1262177}{\nolinkurl{10.1109/TPAMI.2004.1262177}}.

\bibitem{farhi2014qaoa}
E.~Farhi, J.~Goldstone, and S.~Gutmann, ``A quantum approximate optimization
  algorithm,'' arXiv:1411.4028 [quant-ph], 2014.

\bibitem{cerezo2021vqa}
M.~Cerezo, A.~Arrasmith, R.~Babbush, S.~C. Benjamin, S.~Endo, K.~Fujii, J.~R.
  McClean, K.~Mitarai, X.~Yuan, L.~Cincio, and P.~J. Coles, ``Variational
  quantum algorithms,'' \emph{Nature Reviews Physics}, vol.~3, no.~9, pp.
  625--644, 2021, doi:
  \href{https://doi.org/10.1038/s42254-021-00348-9}{\nolinkurl{10.1038/s42254-021-00348-9}}.

\bibitem{cvar2020}
P.~K. Barkoutsos, G.~Nannicini, A.~Robert, I.~Tavernelli, and S.~Woerner,
  ``Improving variational quantum optimization using {CVaR},'' \emph{Quantum},
  vol.~4, p. 256, 2020, doi:
  \href{https://doi.org/10.22331/q-2020-04-20-256}{\nolinkurl{10.22331/q-2020-04-20-256}}.

\bibitem{kourousias2026jinst}
G.~Kourousias, S.~Carrato, F.~Guzzi, M.~Bill{\`e}, A.~Hafner, F.~Bill{\`e},
  R.~Pugliese, and P.~Dimitrakis, ``Quantum-native formulations for {C}omputed
  {T}omography: reconstruction, denoising and {QUBO}-based segmentation,''
  \emph{Journal of Instrumentation}, vol.~21, no.~06, p. C06008, Jun. 2026,
  doi:
  \href{https://doi.org/10.1088/1748-0221/21/06/C06008}{\nolinkurl{10.1088/1748-0221/21/06/C06008}}.

\bibitem{QuantumProvenanceDataset}
G.~Kourousias, S.~Carrato, R.~Pugliese, A.~Hafner, F.~Guzzi, M.~Bill{\`e},
  F.~Bill{\`e}, and P.~Dimitrakis, ``Dataset of sufficient quantum
  provenance,'' Enodo, 2026, doi:
  \href{https://doi.org/10.34965/enodo.i61153}{\nolinkurl{10.34965/enodo.i61153}}.

\bibitem{montanaro2016}
A.~Montanaro, ``Quantum algorithms: An overview,'' \emph{npj Quantum
  Information}, vol.~2, p. 15023, 2016, doi:
  \href{https://doi.org/10.1038/npjqi.2015.23}{\nolinkurl{10.1038/npjqi.2015.23}}.

\bibitem{grover1996}
L.~K. Grover, ``A fast quantum mechanical algorithm for database search,'' in
  \emph{Proceedings of the 28th Annual ACM Symposium on Theory of Computing
  (STOC '96)}.\hskip 1em plus 0.5em minus 0.4em\relax ACM, 1996, pp. 212--219,
  doi:
  \href{https://doi.org/10.1145/237814.237866}{\nolinkurl{10.1145/237814.237866}}.

\bibitem{cleve1998}
R.~Cleve, A.~Ekert, C.~Macchiavello, and M.~Mosca, ``Quantum algorithms
  revisited,'' \emph{Proceedings of the Royal Society of London A}, vol. 454,
  no. 1969, pp. 339--354, 1998, doi:
  \href{https://doi.org/10.1098/rspa.1998.0164}{\nolinkurl{10.1098/rspa.1998.0164}}.

\bibitem{shor1997}
P.~W. Shor, ``Polynomial-time algorithms for prime factorization and discrete
  logarithms on a quantum computer,'' \emph{SIAM Journal on Computing},
  vol.~26, no.~5, pp. 1484--1509, 1997, doi:
  \href{https://doi.org/10.1137/S0097539795293172}{\nolinkurl{10.1137/S0097539795293172}}.

\bibitem{hhl2009}
A.~W. Harrow, A.~Hassidim, and S.~Lloyd, ``Quantum algorithm for linear systems
  of equations,'' \emph{Physical Review Letters}, vol. 103, no.~15, p. 150502,
  2009, doi:
  \href{https://doi.org/10.1103/PhysRevLett.103.150502}{\nolinkurl{10.1103/PhysRevLett.103.150502}}.

\bibitem{aaronson2015fineprint}
S.~Aaronson, ``Read the fine print,'' \emph{Nature Physics}, vol.~11, no.~4,
  pp. 291--293, 2015, doi:
  \href{https://doi.org/10.1038/nphys3272}{\nolinkurl{10.1038/nphys3272}}.

\bibitem{gilyen2019qsvt}
A.~Gily{\'e}n, Y.~Su, G.~H. Low, and N.~Wiebe, ``Quantum singular value
  transformation and beyond: exponential improvements for quantum matrix
  arithmetics,'' in \emph{Proceedings of the 51st Annual ACM SIGACT Symposium
  on Theory of Computing}.\hskip 1em plus 0.5em minus 0.4em\relax ACM, 2019,
  pp. 193--204, doi:
  \href{https://doi.org/10.1145/3313276.3316366}{\nolinkurl{10.1145/3313276.3316366}}.

\bibitem{martyn2021qsvt}
J.~M. Martyn, Z.~M. Rossi, A.~K. Tan, and I.~L. Chuang, ``Grand unification of
  quantum algorithms,'' \emph{PRX Quantum}, vol.~2, no.~4, p. 040203, 2021,
  doi:
  \href{https://doi.org/10.1103/PRXQuantum.2.040203}{\nolinkurl{10.1103/PRXQuantum.2.040203}}.

\bibitem{tang2019}
\BIBentryALTinterwordspacing
E.~Tang, ``A quantum-inspired classical algorithm for recommendation systems,''
  in \emph{Proceedings of the 51st Annual ACM SIGACT Symposium on Theory of
  Computing}, 2019, pp. 217--228, doi:
  \href{https://doi.org/10.1145/3313276.3316310}{\nolinkurl{10.1145/3313276.3316310}}.
  [Online]. Available: \url{https://arxiv.org/abs/1807.04271}
\BIBentrySTDinterwordspacing

\bibitem{albash2018}
T.~Albash and D.~A. Lidar, ``Adiabatic quantum computation,'' \emph{Reviews of
  Modern Physics}, vol.~90, no.~1, p. 015002, 2018, doi:
  \href{https://doi.org/10.1103/RevModPhys.90.015002}{\nolinkurl{10.1103/RevModPhys.90.015002}}.

\bibitem{kadowaki1998}
T.~Kadowaki and H.~Nishimori, ``Quantum annealing in the transverse {I}sing
  model,'' \emph{Physical Review E}, vol.~58, no.~5, pp. 5355--5363, 1998, doi:
  \href{https://doi.org/10.1103/PhysRevE.58.5355}{\nolinkurl{10.1103/PhysRevE.58.5355}}.

\bibitem{hauke2020}
P.~Hauke, H.~G. Katzgraber, W.~Lechner, H.~Nishimori, and W.~D. Oliver,
  ``Perspectives of quantum annealing: methods and implementations,''
  \emph{Reports on Progress in Physics}, vol.~83, no.~5, p. 054401, 2020, doi:
  \href{https://doi.org/10.1088/1361-6633/ab85b8}{\nolinkurl{10.1088/1361-6633/ab85b8}}.

\bibitem{dalzell2025endtoend}
A.~M. Dalzell, S.~McArdle, M.~Berta, P.~Bienias, C.-F. Chen, A.~Gily{\'e}n,
  C.~T. Hann, M.~J. Kastoryano, E.~T. Khabiboulline, A.~Kubica, G.~Salton,
  S.~Wang, and F.~G. S.~L. Brand{\~a}o, \emph{Quantum Algorithms: A Survey of
  Applications and End-to-end Complexities}.\hskip 1em plus 0.5em minus
  0.4em\relax Cambridge University Press, 2025, doi:
  \href{https://doi.org/10.1017/9781009639651}{\nolinkurl{10.1017/9781009639651}}.

\bibitem{jaques2025qram}
S.~Jaques and A.~G. Rattew, ``{QRAM}: A survey and critique,'' \emph{Quantum},
  vol.~9, p. 1922, 2025, doi:
  \href{https://doi.org/10.22331/q-2025-12-02-1922}{\nolinkurl{10.22331/q-2025-12-02-1922}}.

\bibitem{zhao2026oraclesketching}
\BIBentryALTinterwordspacing
H.~Zhao, A.~Zlokapa, H.~Neven, R.~Babbush, J.~Preskill, J.~R. McClean, and
  H.-Y. Huang, ``Exponential quantum advantage in processing massive classical
  data,'' arXiv:2604.07639v2, 2026, doi:
  \href{https://doi.org/10.48550/arXiv.2604.07639}{\nolinkurl{10.48550/arXiv.2604.07639}}.
  [Online]. Available: \url{https://arxiv.org/abs/2604.07639v2}
\BIBentrySTDinterwordspacing

\bibitem{markidis2026oraclesketching}
\BIBentryALTinterwordspacing
S.~Markidis, ``Practical considerations for processing massive classical data
  via quantum oracle sketching,'' arXiv:2609.13549, 2026, doi:
  \href{https://doi.org/10.48550/arXiv.2609.13549}{\nolinkurl{10.48550/arXiv.2609.13549}}.
  [Online]. Available: \url{https://arxiv.org/abs/2609.13549}
\BIBentrySTDinterwordspacing

\bibitem{hashim2025benchmarking}
A.~Hashim, L.~B. Nguyen, N.~Goss, B.~Marinelli, R.~K. Naik, T.~Chistolini,
  J.~Hines, J.~P. Marceaux, Y.~Kim, P.~Gokhale, T.~Tomesh, S.~Chen, L.~Jiang,
  S.~Ferracin, K.~Rudinger, T.~Proctor, K.~C. Young, I.~Siddiqi, and
  R.~Blume-Kohout, ``Practical introduction to benchmarking and
  characterization of quantum computers,'' \emph{PRX Quantum}, vol.~6, no.~3,
  p. 030202, 2025, doi:
  \href{https://doi.org/10.1103/PRXQuantum.6.030202}{\nolinkurl{10.1103/PRXQuantum.6.030202}}.

\bibitem{gidney2021factoring}
C.~Gidney and M.~Eker{\aa}, ``How to factor 2048 bit {RSA} integers in 8 hours
  using 20 million noisy qubits,'' \emph{Quantum}, vol.~5, p. 433, 2021, doi:
  \href{https://doi.org/10.22331/q-2021-04-15-433}{\nolinkurl{10.22331/q-2021-04-15-433}}.

\bibitem{ImagingExpositionDataset}
G.~Kourousias, S.~Carrato, and R.~Pugliese, ``Dataset of {Quantum Computing
  through Computational Imaging}: Supplementary code, data, and web
  companion,'' Elettra Sincrotrone Trieste, data set, 2026, doi:
  \href{https://doi.org/10.34965/enodo.i61162}{\nolinkurl{10.34965/enodo.i61162}}.

\end{thebibliography}
